\documentclass[twocolumn]{aastex631}

\newcommand{\gsrg}{{\tt SRGonG}} 
\newcommand{\chandra}{{\sl Chandra}}

\newcommand{\vorcell}{\mathcal{V}}
\newcommand{\fov}{\mathcal{F}}
\newcommand{\Ksrc}{K}
\newcommand{\llik}{\mathcal{L}}
\newcommand{\proflik}{\mathcal{L}_{\rm profile}}
\newcommand{\mpar}{m_{\rm seg}}

\newcommand{\obsph}{n}

\newcommand{\Sk}{\mathcal{S}}
\newcommand{\Sroi}{\mathbf{S}}
\newcommand{\Data}{\mathbf{X}}
\newcommand{\xph}{\mathbf{x}}
\newcommand{\intx}[1]{\lambda(\xph_{#1})}

\newcommand{\intk}[1]{{\lambda_{#1}}}
\newcommand{\alllam}{\mathbf{\lambda}}
\newcommand{\fulllam}{\mathbf{\Lambda}}
\newcommand{\parasz}{\beta}
\newcommand{\parasnr}{\sigma}
\newcommand{\Nroi}[1]{{{\rm Num}(\Sk_{#1})}}
\newcommand{\Aroi}[1]{{{\rm Area}(\Sk_{#1})}}
\newcommand{\Avor}[1]{{{\rm Area}(\vorcell_{#1})}}
\newcommand{\rmprb}{{\sl f}}
\newcommand{\iid}{\buildrel{\rm iid}\over{\sim}}
\newcommand{\numgrid}{m_{\rm grid}}
\newcommand{\numsubgraph}{m_{\rm graph}}
\newcommand{\numnn}{m_{\rm nn}}
\newcommand{\vorthr}{m_{\rm vorthr}}
\newcommand{\strata}{m_{\rm R}}

\newcommand{\name}{{\gsrg}}
\newcommand{\updatebf}[1]{{#1}}
\usepackage{bm}
\usepackage{bbm}
\usepackage{amsmath}
\usepackage{subfigure}
\usepackage[ruled,vlined]{algorithm2e}
\usepackage{dsfont}
\usepackage{mathtools}

\submitjournal{AJ}

\shorttitle{Identifying diffuse structures}
\shortauthors{Fan et al.}

\begin{document}

\title{Identifying diffuse spatial structures in high-energy photon lists}

\correspondingauthor{Thomas C. M. Lee}
\email{tcmlee@ucdavis.edu}

\author{Minjie Fan}
\affiliation{Department of Statistics,  
University of California, Davis,  \\
One Shields Avenue,  \\
Davis, CA 95616, USA}

\author{Jue Wang}
\affiliation{Department of Statistics,  
University of California, Davis,  \\
One Shields Avenue,  \\
Davis, CA 95616, USA}

\author[0000-0002-3869-7996]{Vinay L.\ Kashyap}
\affiliation{Center for Astrophysics $|$ Harvard \& Smithsonian,   \\
60 Garden Street,\\
Cambridge, MA 02138, USA}

\author[0000-0001-7067-405X]{Thomas C.\ M.\ Lee}
\affiliation{Department of Statistics,  
University of California, Davis,  \\
One Shields Avenue,  \\
Davis, CA 95616, USA}

\author[0000-0002-0816-331X]{David A.\ van Dyk}
\affiliation{Statistics Section, Department of Mathematics, 
Imperial College London,  \\
180 Queen's Gate,  \\
London, SW7 2AZ, UK}

\author{Andreas Zezas}
\affiliation{Physics Department, 
deUniversity of Crete, \\
P. O. Box 2208, GF-710 03,  \\
Heraklion, Crete, Greece}



\begin{abstract}
Data from high-energy observations are usually obtained as lists of photon events.  A common analysis task for such data is to identify whether diffuse emission exists, and to estimate its surface brightness, even in the presence of point sources that may be superposed.
We have developed a novel 
non-parametric event list segmentation algorithm
to divide up the field of view into distinct emission components.  We use photon location data directly, without binning them into an image. 
We
first construct a graph from the Voronoi tessellation of the observed photon locations and then grow segments using a new adaptation of seeded region growing, that we call 
{{\it Seeded Region Growing on Graph}, after which the overall method is named \gsrg.} Starting with a set of seed locations, this results
in an over-segmented dataset, which \name\ then coalesces using a greedy algorithm where adjacent segments are merged to minimize a model comparison statistic;
{we use} 
the Bayesian Information Criterion. 
{Using \name\ we are able}
to identify point-like and diffuse extended sources in the data with equal facility.  We validate \name\ using simulations, demonstrating that it is capable of discerning irregularly shaped low surface-brightness emission structures as well as point-like sources with strengths comparable to that seen in typical X-ray data. We demonstrate \name's use on the \chandra\ data of the Antennae galaxies, and show that it segments the complex structures appropriately.
\end{abstract}




\section{Introduction} \label{sec:intro}

A challenge often encountered in high-energy astronomical analysis is that the images are photon starved and sparse, and contain many `empty' pixels.  Unlike photon-rich images encountered at longer wavelengths, complex features in X-ray and $\gamma$-ray data are difficult to recognize, characterize, and analyze.  Working directly with Poisson distributed photon counts, while simultaneously separating out the contribution of the background, is a difficult process, especially when trying to detect faint non-uniform emission, or separating faint point sources from larger scale diffuse emission.  Finding the boundaries of extended structures is thus a challenging problem.
Such complex structures are common in high-energy astronomical images and include, for example, shock fronts, knots in supernova remnants, regions of diffuse emission in galaxies, point sources embedded in diffuse emission or conglomerates of point sources, entire galaxies or groups/clusters of galaxies, jets, or star forming regions which appear to be extended in the X-ray band even with intermediate resolution ($\lesssim0.5\arcmin$) X-ray telescopes.

The analysis of extended X-ray sources is critical for several areas of astrophysics.  The spatial scales of extended emission contain information regarding the physical processes that lead to their formation, while their boundaries are often determined by their physical environments.  Therefore, identifying the boundary of these regions in a data-driven, rather than a model-driven, fashion is necessary for the scientifically valuable results.  In such cases, a primary goal of the researcher is to segment the image into regions with similar properties and to analyze each segment individually.

Multiscale methods like wavelets \citep{starck:02} for point source detection has been efficiently implemented for X-ray images 
\citep{Freeman-et-al02},
and matched-filter techniques have been successfully used to detect galaxy clusters in ROSAT data \citep{Vikhlinin+1998},
but extended structures remain difficult to find and characterize in these low counts Poisson data.
Other techniques are generally optimized for high S/N images: they apply adaptive binning, or set S/N thresholds to smoothed images with point sources removed \cite[e.g.,][]{Sanders2001,Sanders2006}; adapt methods developed for the analysis of cosmic microwave background images \citep[\updatebf{e.g.,}][]{Bobin+2016}; limit themselves to restrictive assumptions like modeling a combination of point sources
\citep[(E)BASCS;][]{Jones-15,Meyer-21}, or require spectral model similarity across the field of view \citep{Picquenot+2019,2021A&A...646A..82P}.
Currently, most astronomical images with complex structures  that are processed for public display 
use some form of flux-non-conserving adaptive smoothing \citep{Ebeling-et-al06}.
This approach is inadequate for scientific analysis.  Previous efforts at extended source detection using Voronoi tessellation techniques have been limited by computational cost and the imposition of global thresholding schemes \citep[e.g., {\tt vtpdetect},][]{Ebeling-Wiedenmann93}.  Methods akin to seeded region growing 
(cf.\ {\tt SrcExtractor}, \citealt{BertinArnouts96}; {\tt NoiseChisel}, \citealt{2015ApJS..220....1A})
and Machine Learning
(e.g., {\tt Morpheus}, \citealt{2020ApJS..248...20H}; {\tt Mask R-CNN}, \citealt{FARIAS2020100420}; {\tt galmask}, \citealt{2022RNAAS...6..128G})
have been used for the identification of features in optical images of galaxies.
However, the Poisson nature and the sparsity of the X-ray data requires statistically better targeted methods.

Here we develop a new method that combines aspects of Voronoi tessellation with region growing by using neighbor similarity clustering. The method can be applied to X-ray data and provides both separation between different structures in a complex image and well-defined apertures to perform photometry. 
We describe the statistical model that underlies the method in Section~\ref{sec:method}, and specific implementation details including computational methods in Section~\ref{sec:implement}.  We carry out several simulations to test the limits of applicability of the algorithm in Section~\ref{sec:simulation}, and apply it to {\sl Chandra} data of the Antennae 
galaxies in Section~\ref{sec:application}.  We discuss how and when the algorithm may be best used in \ref{sec:discuss}, and summarize our work in Section~\ref{sec:summary}.

\section{Statistical Methodology} \label{sec:method}

We have developed a method that iteratively 
aggregates contiguous sets of photons into distinct regions 
based on 
similarity of their surface brightness.
We employ a likelihood based method to obtain a piece-wise constant estimate of the surface brightness across the image; the likelihood function is derived in  Section~\ref{sec:statmodel}.  The method starts with the high-resolution segmentation of the spatial distribution of the events based on the Voronoi cells described in Section~\ref{sec:voronoi} and combines segments by optimizing the Bayesian Information Criterion (BIC) given in Section~\ref{sec:BIC}.
Table~\ref{tab:notation} provides a glossary of our notation.

\begin{table*}[ht]
    \centering
    \caption{Glossary of variables and notation.}
    \begin{tabular}{l l}
    \hline\hline
    {\bf Notation} & {\bf Description} \\
    \hline
    $\iid$ & Independent and identically distributed \\
    $\hat{\zeta}$ & Estimate of a generic parameter $\zeta$ \\
    $\fov$ & Field of view, a bounded domain in $\mathbb{R}^2$ that contains the observed photons  \\
    $\obsph$ & Number of observed photons \\ 
    $\Data = \{\xph_1, \ldots, \xph_\obsph \}$ & Observed location of the $\obsph$ photons, denoting their
    (sky) coordinates \\
    $\Ksrc$ & Number of segments within $\fov$ \\
    $\Sroi = \{\Sk_1, \ldots, {\Sk_\Ksrc}\}$ & Partition of $\fov$, where $\Sk_{k}$ is the domain for segmented region $k$ \\
    $\Nroi{k}$ & Number of photons observed in segment $\Sk_k$ \\
    $\Aroi{k}$ & Area of segment $\Sk_k$ \\
    $\mpar$ & Number of free parameters per segment \\
    $\intx{}$ & Poisson intensity at location $\xph$ \\
    $\bm{\alllam}=\{\intk{1},\ldots,\intk{\Ksrc}\}$ & Collection of the Poisson intensities over each of the segments \\
    $\fulllam$ & Total integrated intensity of  $\intx{}$ over $\fov$ \\
    $\vorcell_i$ & Voronoi cell defined by photon $i$ \\
    $\Avor{i}$ & Area of Voronoi cell $\vorcell_i$ \\
    $\widehat\alllam_i^{{\rm \vorcell}_i}$ & The Voronoi estimator of the Poisson intensity across Voronoi cell $i$, with $\widehat\alllam_i^{{\rm \vorcell}_i} =1/\Avor{i}$ \\
    $\rmprb(\Data, \obsph)$ & Joint probability mass/density function of
    the observed number of photons and their locations \\
    $\llik(\Ksrc,\Sroi,\bm{\alllam}\ | \ \Data)$ & Log-likelihood of the model$^a$ \\
    $\proflik(\Ksrc,\Sroi\ | \ \Data)$ & Profile log-likelihood$^a$ obtained by replacing $\alllam$ in $\llik(\Ksrc,\Sroi,\alllam\ | \ \xph)$ with its estimate $\hat{\alllam}$ \\
    ${\rm BIC}(\Ksrc)$ & Bayesian Information Criterion as a function of the number of segments, $\Ksrc$ \\
    $\numgrid$ & {In seed specification, the } number of grid points used in a regularly spaced grid \\
    $\numsubgraph$ & Number of photons included in initial subgraph for each seed \\
    $\numnn$ & Number of nearest neighbors over which local maxima are searched for to specify seeds \\
    $\strata$ & Number of strata used during Voronoi-area stratified sampling to specify seeds \\
    $\vorthr$ & Minimum number of photons required
    {for} a Voronoi-area derived seed
    {to be} accepted \\
    \hline
    \multicolumn{2}{l}{$a:$ Notation is reversed, by convention, for $\llik(b|a)$ compared to conditional probabilities; e.g., } \\ [-5pt]
    \multicolumn{2}{l}{\quad \ ${\rm p}(a|b)$ represents the probability of $a$ given $b$.}
    \end{tabular}
    \label{tab:notation}
\end{table*}

\subsection{Statistical Model}
\label{sec:statmodel}

We consider an event list composed of $\obsph$ 
photons observed in a bounded domain that defines the field of view, $\fov \subset \mathbb{R}^2$. Ignoring instrumental pixelization, we model the set of sky coordinates for the $\obsph$ photons, 
\begin{equation}
    \Data=\{\xph_1,\ldots,\xph_\obsph\} \,,
\end{equation}
via an inhomogeneous Poisson process with intensity function $\intx{} \ge 0$.
The intensity function must be integrable over $\fov$, i.e.,  $\fulllam = \int_\fov\intx{} d\xph$ must be finite. For simplicity, we assume that the intensity function is piece-wise constant. Specifically, we assume we can partition $\fov$ into $K$ segments, denoted $\Sroi=(\Sk_1, \ldots, \Sk_{\Ksrc})$, where the Poisson intensity is constant on each $\Sk_k$. Since $\Sroi$ partitions $\fov$, the $\Sk_k$ together cover $\fov$ and each pair is disjoint.  For a given set of non-negative intensities,
$\bm{\alllam}=\{\intk{1},\ldots,\intk{\Ksrc}\}$,
we can then express the intensity function as
\begin{equation}
    \intx{} =\sum_{k=1}^{\Ksrc} \intk{k} \mathbbm{1}_{\Sk_k}(\xph) \,,
    \label{eq:pw-constant}
\end{equation}
where $\mathbbm{1}_{\Sk_k}(\xph)$ is an indicator function that takes value 1 if $\xph \in \Sk_k$
and is otherwise 0. 

A property of the inhomogeneous Poisson process is that the
number of photons, $\Nroi{k}$, recorded in segment $\Sk_k$ with area $\Aroi{k}$
follows a Poisson distribution with mean
\begin{equation}
    \Aroi{k} \cdot \intk{k} = \int_{\Sk_k}\intx{} d\xph,  ~\hbox{ for }~ k=1,\ldots, \Ksrc \,,
\end{equation}
with $\intk{k} \ge 0$. 
Likewise, the total photon count is distributed
$\obsph \sim {\rm Pois}(\fulllam)$. Another property is that, given $\obsph$, the sky coordinates, $\xph_i$ are independent and identically distributed (iid) with (normalized) probability density function $\intx{}/\fulllam$ \citep[e.g.,][]{chiu2013stochastic}. This means that the $\xph_i$ are distinct -- no two photons can have the same recorded coordinates. (The discrete nature of detectors means that occasionally, two photons are recorded with identical coordinates. In this case, we add a very small random scatter, $\sim$10$^{-6}$.) 
Our goal is to estimate the number of segments, $K$, the segments, $\Sroi$, and their respective intensities, $\bm{\alllam}$.

Thus far, we have not discussed sources or background. If the field of view includes multiple separated sources, we expect the piece-wise constant intensity function to capture the intensity peaks associated with point and extended sources. Between these sources (or around a single source) is the background region. If the background intensity is constant across $\fov$ and the source region(s) is/are isolated within the field of view, we expect a single large segment representing the background to encircle the source regions and to extend to the boundary of $\fov$. If the background intensity varies slowly, we might find several large segments that together comprise the background region. In any case, there are segments associated with background and with sources. We do not attempt to classify the segments in this regard. Of course, if there is a single large low-intensity segment encircling smaller higher intensity segments, it is easy enough to identify the background with the large low-intensity segment.  

We are particularly interested in the case where small-scale point-like sources lie within a larger extended source as this is a challenging task for existing methods. We do not distinguish between extended or point sources, and in fact ignore the effect of telescope's point spread function, assuming that it is small compared to the size of the $\fov$.  Our method is agnostic by design to the sizes of individual structures, and is thus capable of isolating sources at all scales.

To derive the likelihood function, recall  $\obsph \sim {\rm Pois}(\fulllam)$ and given $\obsph$ the $\xph_i \iid \intx{}/\fulllam$.
Thus, their joint probability mass/density function under the inhomogeneous Poisson process is 
\begin{eqnarray}
\rmprb(\xph_1, \ldots, \xph_n, n) & = & \rmprb(\Data \mid n) \cdot \rmprb(n) \nonumber \\
& = & \frac{1}{\fulllam^n} 
\prod_{i=1}^n \intx{i}  \cdot
 \frac{\exp\left({-\fulllam} \right) \fulllam^n}{n!} \nonumber \\
& = & 
\frac{\exp\left({-\fulllam} \right)}{n!}
\prod_{i=1}^n \intx{i}
\end{eqnarray}
and their log-likelihood function is given by
\begin{multline}
\llik(\Ksrc, \Sroi, \bm{\alllam}\mid \Data, n) 
= \log{\rmprb(\Data, n)} \\
=\sum \limits_{i=1}^n \log{\intx{i}}-\int_\fov\intx{} d\xph-\log{n!} ~~,
\end{multline}
where we write out $\fulllam = \int_\fov\intx{} d\xph$. Replacing $\intx{}$ by the piece-wise constant expression given in (\ref{eq:pw-constant}), we have
\begin{multline}
\llik(\Ksrc, \Sroi, \bm{\alllam} \mid \Data, n) =\\
\sum \limits_{\substack{k=1 \\ \Nroi{k}\neq 0}}^{\Ksrc} 
\kern -13pt \Nroi{k} \log {\intk{k}} 
-\sum_{k=1}^{\Ksrc} \Aroi{k}\intk{k}-\log{n!} ~~.
\label{eq:loglike}
\end{multline}
Recall that
$\Nroi{k}$ and $\Aroi{k}$ denote the number of photons in $\Sk_k$ and the area of $\Sk_k$, respectively. (When $\Nroi{k}=0$, the summand $$\Nroi{k} \log {\intk{k}}$$ is excluded from the first sum in (\ref{eq:loglike}).)

We aim to maximize $\llik$ as a function of $K$, $\Sroi$, and $\bm{\alllam}$ 
to obtain their maximum likelihood estimates.
For fixed $K$ and $\Sroi$, $\llik$ is maximized as a function of $\bm{\alllam}$ by
\begin{equation}
    \widehat{\lambda}_k =\Nroi{k}/\Aroi{k} ~\hbox{ for }~ k=1,\ldots, \Ksrc \,.
\end{equation}
Plugging the $\widehat{\lambda}_k$ into $\llik$, we obtain the profile log-likelihood of $K$ and $\Sroi$, i.e., 
\begin{multline}\label{eqn:prof_log_lik}
\proflik(\Ksrc, \Sroi \mid \Data, n)=\\
\sum \limits_{\substack{k=1 \\ \Nroi{k}\neq 0}}^{\Ksrc} 
\kern -10pt \Nroi{k} \log \left(\frac{\Nroi{k}}{\Aroi{k}}\right)-n-\log{n!}.
\end{multline}
The same profile log-likelihood can be derived by 
modeling the data as a mixture of uniform distributions. \citet{Allard-97} considered a special case with
a uniform background with a contiguous extended source superposed.

To estimate $\Sroi$, we first deploy a (greedy) algorithm that finds an optimal segmentation, $\widehat\Sroi(\Ksrc) =
{\operatorname{arg\, max}} \proflik(\Ksrc, \Sroi)$, for fixed $\Ksrc$, as described in Sections~\ref{sec:voronoi} and refined in Section~\ref{sec:implement}. In Section~\ref{sec:BIC}, we introduce a penalized version of $\proflik$ that we maximize over $\Ksrc$ to obtain final estimates of the number of segments, $\widehat\Ksrc$, and thereby of the segments themselves, $\widehat\Sroi(\widehat\Ksrc)$.

\subsection{Estimating $\Sroi$ via Voronoi Tessellation}
\label{sec:voronoi}

\begin{figure*}[hbt!]
    \centering
    \includegraphics[width=0.32\linewidth]{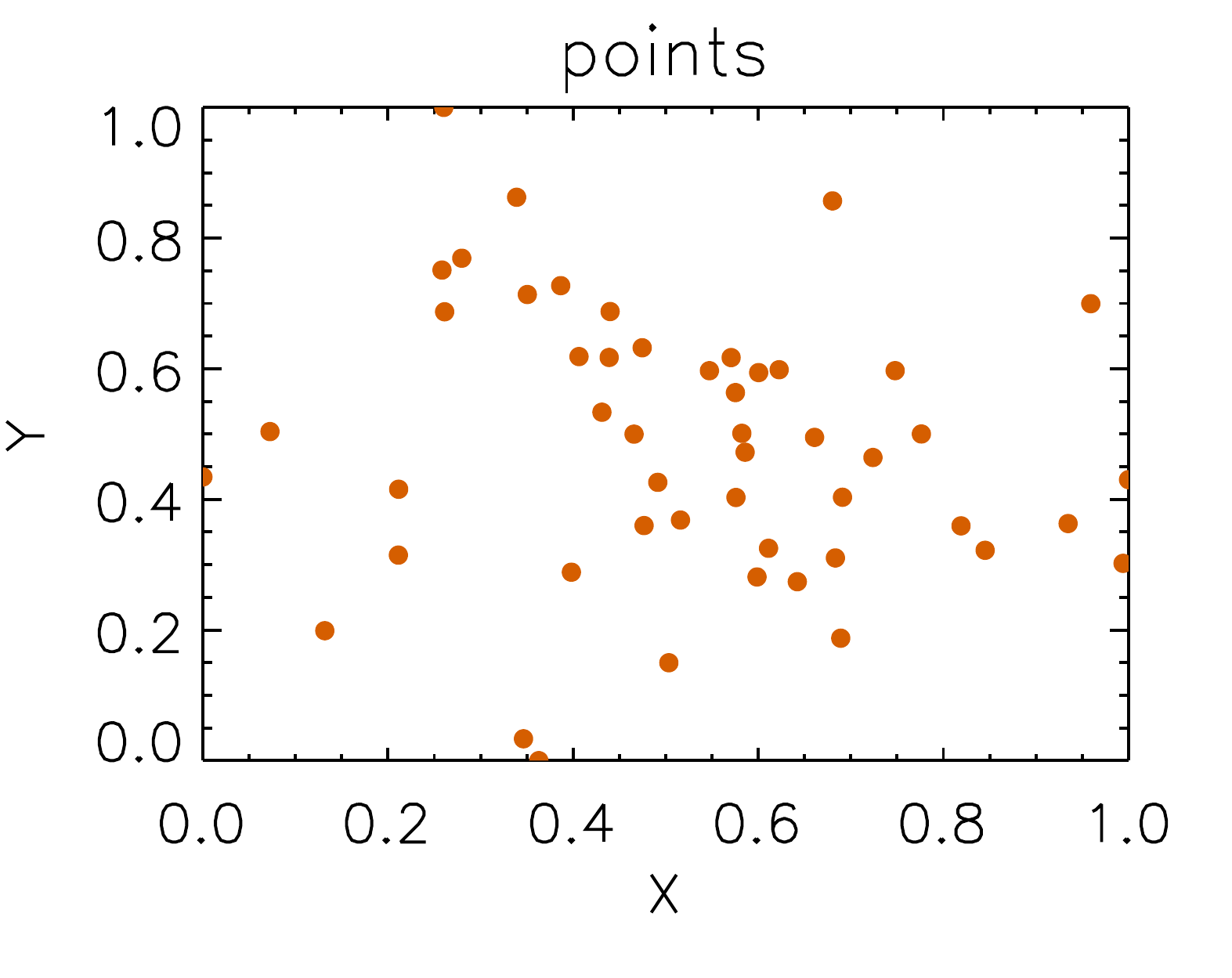}
    \includegraphics[width=0.32\linewidth]{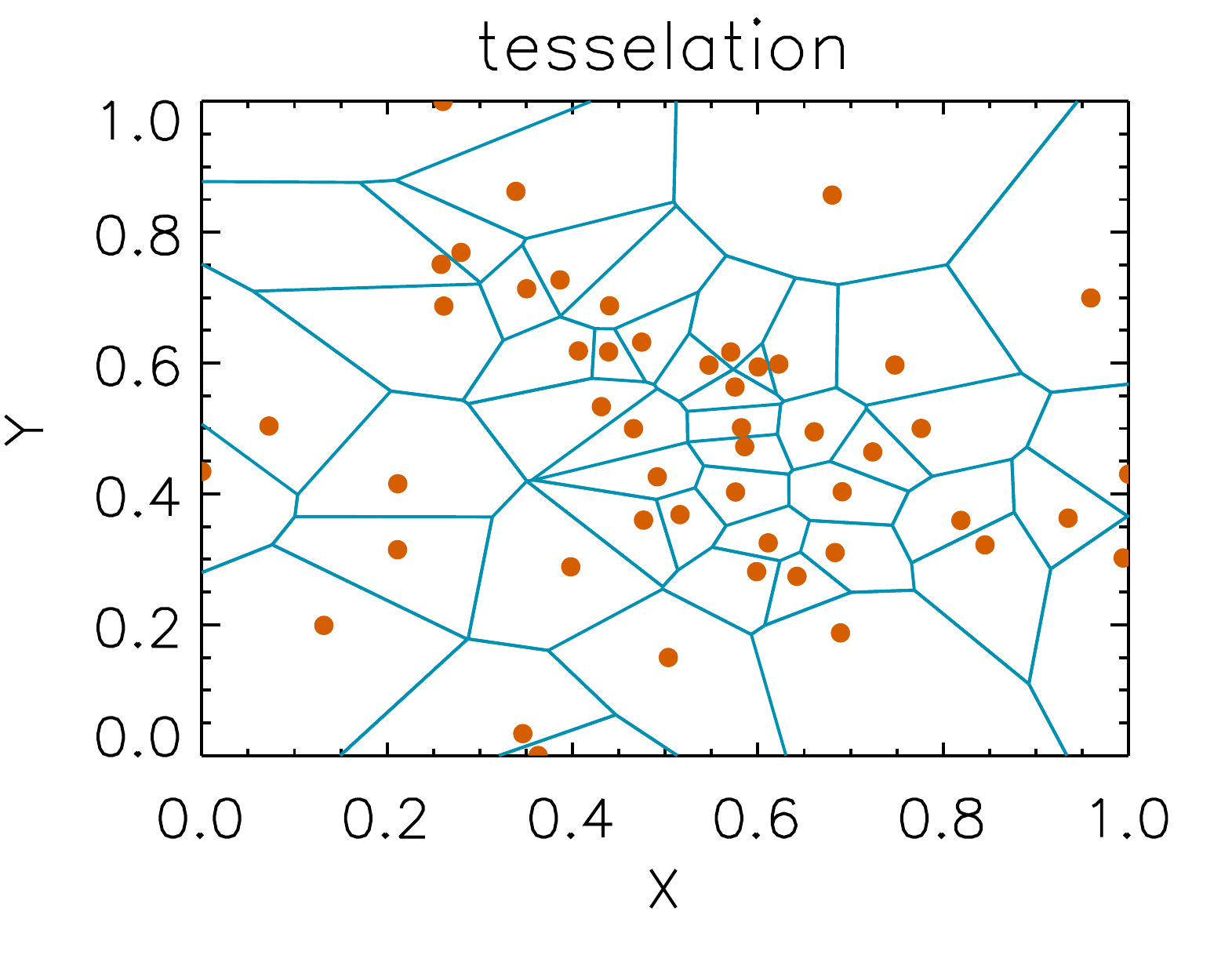}
    \includegraphics[width=0.32\linewidth]{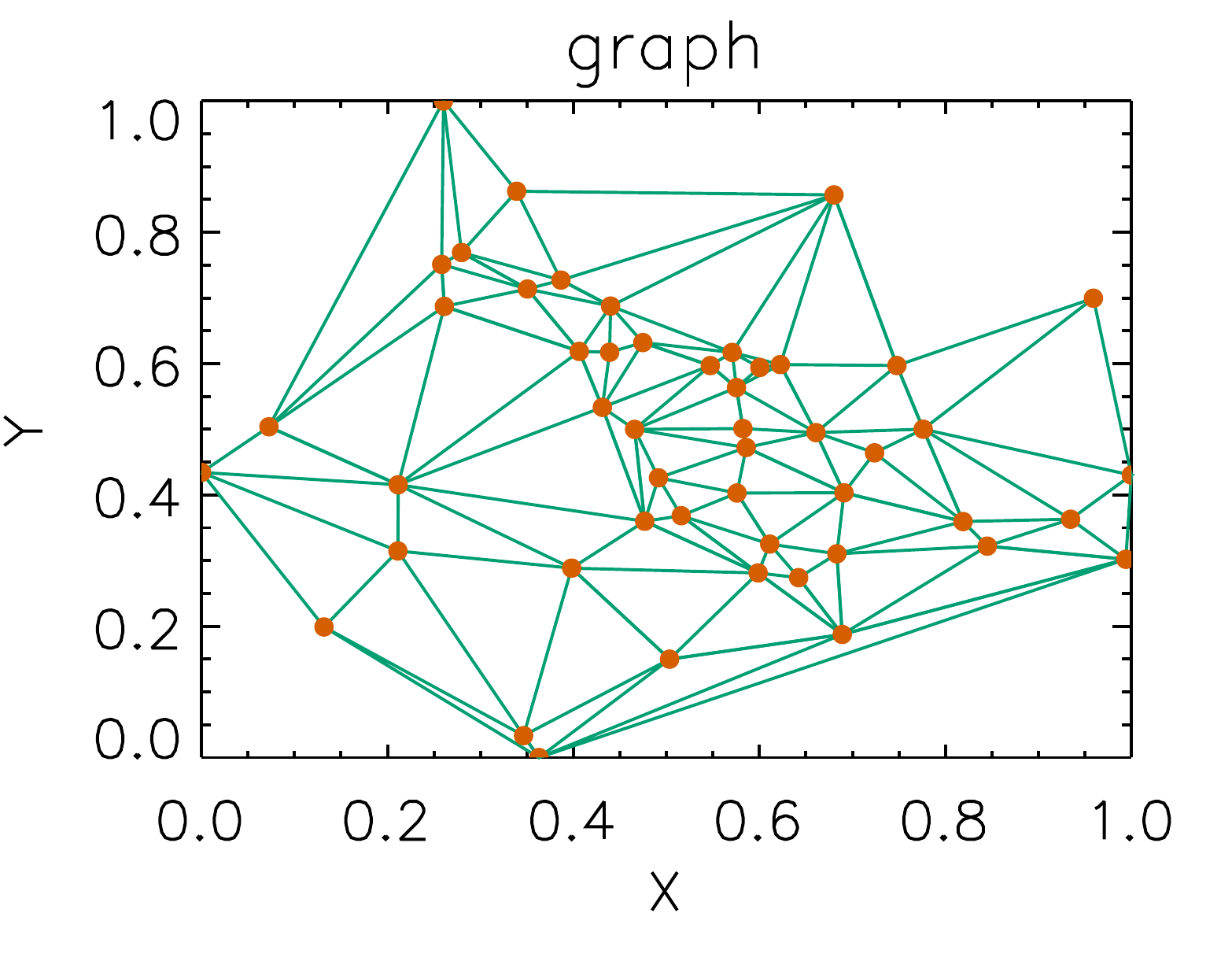}
    \caption{Illustration of tessellation and triangulation of points.  \textbf{Left Panel:} 50 points drawn randomly from a multivariate normal $N(\mathbf{0.5},\mathbf{0.2^2}{\cal{I}})$ are plotted as red dots,
    where ${\cal{I}}$ is a unit matrix.  \textbf{Middle Panel:} The Voronoi cells generated by these points are shown as cyan polygons surrounding each point.  \textbf{Right Panel:} The corresponding graph of Delaunay triangulation is shown as green line segments connecting adjacent points. }
    \label{fig:voronoi_illustration}
\end{figure*}

Obtaining estimates of the segments, $\Sroi$, requires us to constrain the set of possible partitions. For any fixed $\Ksrc$, for example, we can make $\proflik$ arbitrarily large by including a segment that is small enough to contain exactly one photon  and shrinking the segment's area toward zero (since  $\Aroi{k}$ appears in the denominator of (\ref{eqn:prof_log_lik})).  Similarly, any  $\Sk_k$ with $\Nroi{k}=0$ can have arbitrary shape 
since it does not contribute to the profile likelihood. 
Since we cannot estimate the intensity function at a higher resolution than the data, we only consider candidate segments that include at least one photon.

One way to do constrain $\Sroi$ is to only consider candidate segments that consist of the Voronoi cells derived from the Voronoi tessellation of the data, 
or the union of several Voronoi cells. 
The Voronoi tessellation of the observed photons uniquely partitions $\fov$ into $n$ convex cells, denoted $\vorcell_i, ~i=1, \ldots, n$, such that cell $\vorcell_i$ contains exactly one photon, 
say $\xph_i$, and consists of all locations in $\fov$ closer to photon $\xph_i$ than to any other photon. These cells are called Voronoi cells, and $\xph_i$ is called the nucleus of $\vorcell_i$. Figure~\ref{fig:voronoi_illustration} gives an example of the Voronoi tessellation of 50 photons drawn from a normal distribution truncated to the unit square. The photon locations are plotted in the left panel and the Voronoi cells in the middle panel. (We discuss the graph in the right panel in Section~\ref{sec:gsrg}.)

To avoid unclosed Voronoi cells near the border of the field of view, we restrict the tessellation to Voronoi cells whose vertices are all in $\fov$. Based on the Voronoi tessellation, \citet{Barr-10} introduced the Voronoi estimator $\widehat\alllam_i^{{\rm \vorcell}_i}(\xph)=1/\Avor{i}$ for any location $\xph \in \vorcell_i$. They show that under certain conditions, the Voronoi estimator is approximately unbiased for the Poisson intensity $\intx{}$,
and its sampling distribution is approximately the inverse Gamma distribution\footnote{The probability density function of an inverse Gamma
distribution is $\frac{b^{a}}{\Gamma(a)}x^{-a-1}\exp \left( -\frac{b}{x} \right)$, where $x>0$, $a$ and $b$ are the shape and rate parameters, respectively, and $\Gamma(\cdot)$ denotes the Gamma function.
}.

The algorithm that we propose to combine the Voronoi cells to form the segments (by approximately maximizing $\proflik$ for each fixed $\Ksrc$) is detailed in Section~\ref{sec:implement}. When $\Ksrc$ is fixed in advance, this algorithm can be used to estimate $\Sroi$. To fit $\Ksrc$ we use the method in Section~\ref{sec:BIC}.

\subsection{Estimating $\Ksrc$ via the Bayesian Information Criterion}
\label{sec:BIC}

Unfortunately, the number of sources, $\Ksrc$, cannot be reasonably estimated by maximizing the profile likelihood, because (\ref{eqn:prof_log_lik}) increases with  $\Ksrc$ and is thus maximized by $\Ksrc=n$, i.e., with the full set of Voronoi cells\footnote{Since $n$ is fixed, maximizing $\proflik$ is equivalent to maximizing the sum in (\ref{eqn:prof_log_lik}), which can be written as 
$\sum_{i=1}^n \log \hat\lambda(\xph_i)$
where
$\hat\lambda(\xph_i)$ 
is the local optimizer, $\Nroi{} / \Aroi{}$, for the segment containing $\xph_i$. Increasing the number of segments allows for better local optimization of local fluctuations and thus increases $\proflik$. Of course, with too many segments, better fitting of local fluctuations amounts to fitting noise, i.e., over-fitting the data.}. We avoid such over-fitting by adding a term to (\ref{eqn:prof_log_lik})
that suitably penalizes model complexity.  Specifically, we use the so-called Bayesian Information Criterion (BIC), which has been shown to produce statistically consistent results for many model selection problems. For the current problem, the BIC is defined as
\begin{equation}
    {\rm BIC}(\Ksrc)=-2\proflik(\Ksrc, \widehat\Sroi(\Ksrc) \mid \Data, \obsph)+\Ksrc\mpar\log\obsph \,,
    \label{eq:BICdefn}
\end{equation}
where $\mpar$ 
is the number of free/independent parameters per segment, thus $\mpar \Ksrc$ is the total number of free parameters in the model\footnote{Since the shape of the final segment is determined by the first $\Ksrc-1$ segments, a more precise formulation of the total number of parameters in the model is $
\mpar(\Ksrc-1)+1$, where the intensity parameter of the final segment is accounted for by the ``+1''. The difference between this more precise formulation and the one used in (\ref{eq:BICdefn}) is $(1-\mpar)n$, which does not depend on any of the unknown parameters and thus does not affect estimation.}. The BIC 
estimate for $\Ksrc$ is given by
\begin{equation}
    \widehat{\Ksrc} =
    {\operatorname{arg\, min}}~{\rm BIC}(\Ksrc) \,.
    \label{eq:BICval}
\end{equation}
For fixed $\Ksrc$, optimizing BIC is equivalent to optimizing $\proflik$, thus this estimate of $\Sroi$ is equivalent to that described in Section~\ref{sec:voronoi}.

Unfortunately, because we are using a non-parametric model, $\mpar$ is not well-defined. Following \citet{Aue-11}, we set $\mpar$ by approximating the model by a parametric one using an assumed specific shape for the segments. 
When the segments are close to ellipse-shaped, for example, $\mpar=6$ to account for the coordinates of the center, lengths of the two axes, orientation, and intensity.
When the segments are close to circular, $\mpar$ is reduced to $4$. 
Another possibility is to simply set $\mpar = 1$ and the number of model parameters to the number of segments, which to some extent reflects the overall model complexity \citep{Magnussen-06}, but ignores the shapes of the segments. While these parametric approximations allow us to assign a reasonable value to $\mpar$, the model itself remains non-parametric.

BIC is closely related to the ``fitness function" used in the Bayesian block method \citep{Scargle-13}, where model complexity is penalized via a geometric prior on the number of sources, i.e., 
$p(\Ksrc) = P_0\gamma ^{\Ksrc}$, with $\gamma$ being a tuning parameter and $P_0$  a normalization constant. Setting
$\gamma= 1 / {\obsph}^{\Ksrc\mpar}$ 
makes the fitness function equivalent to the penalty term in BIC.

\section{Algorithm for Combining Voronoi Cells into Segments}
\label{sec:implement}

\subsection{\gsrg: Seeded Region Growing on Graph}
\label{sec:gsrg}

Using the Voronoi cells as building blocks, we start by proposing an algorithm to estimate the segments in $\Sroi$ for fixed $\Ksrc$. The first step is to identify pairs or groups of Voronoi cells that can potentially be combined. We accomplish this via the dual graph of the Voronoi tessellation, known as the Delaunay triangulation. This graph's vertices are the centers of the Voronoi cells, i.e., the photons, and its edges connect pairs of the adjacent Voronoi cells. The right panel of Figure~\ref{fig:voronoi_illustration} depicts the graph derived from the Voronoi cells in the middle panel. We assign vertex $\xph_i$ the value of the Voronoi estimator, denoted by
$\widehat\alllam_i^{{\rm \vorcell}_i}$, i.e., an
estimate of the intensity in Voronoi cell $\vorcell_i$. Using the graph constructed by the Delaunay triangulation, the problem of estimating the $\Sk_k$ can be naturally translated to the problem of graph segmentation, i.e., partitioning the graph into subgraphs such that the Voronoi cells therein form a single segment, $\Sk_k$. Thus, each subgraph/$\Sk_k$ is formed by a set of vertices/photons connected by a collection of edges in the full graph. Since we assume that the intensity function is piece-wise constant, all the vertices in each subgraph should share similar values of
$\widehat\alllam_i^{{\rm \vorcell}_i}$.

Unfortunately, finding $\Sroi$ to maximize (\ref{eqn:prof_log_lik}) for fixed $\Ksrc$ remains an intractable combinatorial optimization problem even when confined to combinations of Voronoi cells. A distinct advantage of representing the problem as graph segmentation is that we implicitly impose an additional constraint that each $\Sk_k$ is a subgraph. In this way, traditional image segmentation\footnote{Image segmentation is the process of separating an image into a number of regions such that each region is composed of connected pixels with similar characteristics, such as similar pixel values.} methods can be adapted and used to segment the graph. In particular, we propose the Seeded Region Growing on Graph (\gsrg) method, which is similar to the original Seeded Region Growing (SRG) method used for images except that the concept of ``neighbors'' is determined by the edges of the graph instead of neighboring pixels. The original SRG is proposed in \citet{Adams-94} and is extended to several variants to deal with more complicated cases in \citet{Fan-14}. \gsrg\ starts by identifying, either manually or automatically, a set of initial seeds from the graph.  Each seed can be a single vertex/photon or a seeding subgraph, i.e., a set of connected vertices/photons. 

For the moment,
{we present a simplified version of \gsrg\ that requires}  
a perfect set of seeds, i.e., a set with exactly one seed in each $\hat \Sk_k$. Recall that $\Ksrc$ is fixed, thus initially we assume $\Ksrc$ seeds.
The details of seed specification in more realistic settings are described in Section~\ref{sec:seed_spec} and
{the full version of \gsrg\ (which requires an extra step to merge segments and estimate $\Ksrc$)
}
is detailed in Section~\ref{sec:subgraph_merge}.  

\gsrg\ grows the seeds into subgraphs by successively adding neighboring vertices to them. More specifically, at each iteration, the method selects a pair that consists of a growing subgraph, $S$, and one of its unassigned neighboring vertices, $i$, such that 
\begin{equation}\label{eqn: min_criterion}
\delta(i, S)=\left \lvert \log
\widehat\alllam_i^{{\rm \vorcell}_i} -
\log\{\Nroi{}/\Aroi{}\} \right \rvert
\end{equation}
is minimized.
This criterion compares the logarithm of the estimated intensities of the subgraph and the neighboring vertex because $\proflik$, which we aim to optimize, combines the segment-specific intensity estimates on the log scale.  The vertex in the pair with the smallest difference is added to the corresponding subgraph.
{This process} finishes when all the vertices of the full graph are assigned to exactly one subgraph. The Voronoi cells contained in the subgraphs give the final segmentation of $\fov$, i.e., $\widehat\Sroi(\Ksrc)$, for pre-specified $\Ksrc$.

In practice, we save the index, $i$, of the neighboring Voronoi cell that minimizes $\delta(i, S)$ for each growing subgraph, $\Sk_k$, at each iteration. This reduces the time complexity of the method to be linear in terms of the number of photons.

\subsection{Seed Specification}
\label{sec:seed_spec}

Since \gsrg\ begins by building out regions starting from a specified set of seeds, the number and location of the seeds are important considerations. As discussed in Section~\ref{sec:gsrg}, we would ideally have exactly one seed within each $\hat\Sk_k$. Unfortunately, this is not feasible in practice. A brute force solution is to over-specify the seed set to the extreme, by setting every photon location to be a seed, and devising an algorithm to merge the seeds into segments.  
\updatebf{But merging such a large seed set would be challenging in terms of both computational speed and statistical accuracy (see discussion in Section~\ref{sec:performance_antennae}).  If the field being analyzed is known to have a large number of point sources, or if the scientific question requires focusing on point sources, then running a source detection algorithm first to find all such sources and specifying all of them as seeds will be helpful.  Here we describe three generic strategies to specify smaller initial seed sets.}
These strategies still over-specify the set in that they use a larger number of seeds than the expected number of segments (but less so than setting each photon to be a seed). Thus, after growing the seeds into subgraphs as described in Section~\ref{sec:gsrg}, we require a method to merge the resulting subgraphs into segments; we describe our merging algorithm  in Section~\ref{sec:subgraph_merge}.

\paragraph{Regular grid:} 

This method starts by overlaying a regular grid 
of $\numgrid$ points
onto the field of view, $\fov$. 
For each grid point, we specify a seeding subgraph composed of the $\numsubgraph$ photons closest to the grid point (in terms of the Euclidean distance). 
We typically set   $\numgrid$ and $\numsubgraph$ so that their product is much smaller than $n$ to enable the seeded regions to grow.
Conflicts in the allocation of photons to seeding subgraphs (e.g., when a single photon is among the $\numsubgraph$ closest to two or more grid points) 
are broken by the order of assignment.
The number of the photons, $\numsubgraph$, assigned to each seeding subgraph 
can be increased
to stabilize the initial estimates of the growing subgraph intensities, especially when the contrast
(i.e., the ratio between the intensities of an extended source and the background) is low.
In practice, there is no universally best choice for $\numgrid$ and $\numsubgraph$, as their optimal values depend on factors including the number of observed photons $\obsph$ and the complexity and number of true astronomical sources.
To ensure a sufficient number of photons for the seeding subgraphs, we require $$\numsubgraph{\leq}\frac{\obsph}{\numgrid}\,.$$

Since it is possible for a seed to fall on the boundary between two distinguishable segments and adversely affect subsequent processing, we propose an additional {\sl seed-rejection} step. Specifically, we compare the range of the Voronoi areas for each photon $\Avor{k}|_{\{k=1,\ldots,{\numsubgraph}\}}$ of a seeding subgraph of size  $\numsubgraph$ with the expected empirical $2\sigma$ confidence interval for a homogeneous distribution of photons \citep[][Chapter 4.2]{Moller-94}, i.e.,  
\begin{equation}
    \frac{1}{\widehat\alllam_s} \pm 2\times\frac{0.53}{\widehat\alllam_s}
\end{equation}
\citep[][Chapter 4.2]{Moller-94}, 
where 
\begin{equation}
    \frac{1}{\widehat\alllam_s} = \frac{1}{\numsubgraph}\sum_{k=1}^{\numsubgraph} \Avor{k}
\end{equation}
is the average of Voronoi areas for the photons in the seeding subgraph.  Thus, if the actual range of Voronoi areas $\Avor{k}$ exceeds the expected empirical confidence interval, the seeding subgraph is rejected.

\paragraph{Grid supplemented by local maxima:}

If the regular grid used to generate the seeds is too sparse, some image structures may not be captured in the segmentation. For example, if there is not a grid point sufficiently near a point or extended source, the source may be merged into the background or another source. One remedy is to include additional seeds near the likely locations of sources. Sources induce an elevated intensity over small spatial scales. Thus, locations of high photon density are likely associated with sources.
We propose to identify vertices that are local maxima of the graph constructed by the Delaunay triangulation, in the sense that the vertex value (i.e., $\widehat{\lambda}_i^{{\rm \vorcell }_i}$) is greater than or equal to that of its closest $k$ vertices (including itself), where closeness is measured by the Euclidean distance. 
For each local maxima we find in this way, we include a seeding subgraph composed of its closest $\numsubgraph$ vertices (including itself).

\paragraph{Voronoi-area stratified sampling:} 

More complex schemes, designed to locate seeds over a broader range of surface brightness, can also be devised.  Methods such as Otsu's thresholding \citep{Otsu-79} can also be used to specify seeds or seeding subgraphs for point-like or localized extended sources.  As we discuss in Section~\ref{sec:performance_antennae}, the grid supplemented by local maxima is adequate to identify structures that exist at a large variety of scales in astronomical data.  Here, as an example case, we describe a
third
method, which selects seeds via stratified sampling of the photons, with strata determined by the areas of the Voronoi cells, $\Avor{i}$.  Specifically, the photons are divided into $\strata$ strata bounded by equally spaced quantiles of the distribution of $\Avor{i}$. 
The number of strata depends on the sample size, but we typically use $\strata\approx{10-20}$.
Clumps of near-neighbor photons within each stratum are put together (see Appendix~\ref{sec:percolate})
into a set of labeled groups such that spatially nearby photons within a given stratum are all assigned the same label.  If a given label is assigned to fewer than $\vorthr$ photons (typically $\vorthr=5$), then the photons with this labels are discarded for the purpose of seed specification; otherwise the central photon\footnote{Consider the set, $L$, of photons with a given label.  For each photon $l\in L$, we calculate $d_l = \sqrt{\Avor{l}} + \sum_{k{\in}L}~d_{lk}$ where $d_{lk}$ is the Euclidean distance between photons $l$ and $k$.  That photon in $L$ with the smallest $\{d_l\}$ is flagged as the central photon among the photons in $L$.  This measure of centrality is better than computing a centroid as it ensures that the seed is guaranteed to be included inside the labeled region even when the region shape is complex, and that the seed is unambiguously assigned to one of the photons.} amongst those with each label is set as a seed.  Subgraphs are then constructed for each retained seed photon in the same manner as described in Section~\ref{sec:gsrg}.

\begin{algorithm*}[t]\label{GSRG_algo}
\DontPrintSemicolon
\KwData{Coordinates of observed photons $\xph_i=(x_{1i}, x_{2i}), i=1, \ldots, n$ in field of view $\fov$.}
\KwResult{Piece-wise constant estimate of intensity function with a segmentation of $\fov$ into regions of constant intensity,  $\widehat\Sroi=(\hat\Sk_0, \ldots, \hat\Sk_{\widehat\Ksrc})$.}
\Begin{
\nl Use Voronoi tessellation to obtain a graph whose vertices are the observed photons with the Voronoi estimators $\widehat{\lambda}_i^{{\rm \vorcell}_i}$ as their values.\;
\nl Using a method in Section \ref{sec:seed_spec}, specify the initial seeds for subgraph growing.\;
\nl Grow seeds into subgraphs that over-segment the entire graph:\;
\While{there are unassigned vertices}{
Select a pair of a growing subgraph $S$ and one of its neighboring vertices $i$ such that $
\delta(i, S)=\left \lvert \log
\widehat\alllam_i^{{\rm \vorcell}_i} -
\log\{\Nroi{}/\Aroi{}\} \right \rvert$ is minimized.\;
Add the vertex in the pair with the smallest difference to the corresponding subgraph.\;} 
\nl Greedily merge subgraphs by minimizing BIC at each merger to obtain a nested sequence of segmentations.\;
\nl Finally, set $\widehat\Ksrc$ and $\widehat\Sroi$ to the values of the nesting level with the smallest BIC. $\widehat\Ksrc$ is the final segmentation of $\fov$.\;
}
\caption{Seeded Region Growing on Graph (\gsrg)}
\end{algorithm*}

\subsection{Subgraph Merging}
\label{sec:subgraph_merge}

Using one of the seed sets of Section~\ref{sec:seed_spec} to grow subgraphs as described in Section~\ref{sec:gsrg}
leads to an over-segmented graph since the number of seeds is invariably more than the predetermined $\Ksrc$ or the $\widehat\Ksrc$ that optimizes BIC.
To merge the subgraphs into segments, we propose a subgraph merging method that aims to minimize BIC.
Similar ideas were used by \citet{Lee-00} and \citet{Peng-11} in image segmentation. Specifically, the subgraph merging method starts by computing the BIC for the over-segmented graph  and then iteratively selects two neighboring subgraphs, merges them, and recomputes BIC. The two merged subgraphs are selected so that their merger gives the largest decrease (or the smallest increase) of BIC among all possible merges (of neighboring subgraphs). In this sense, this is a greedy algorithm.  We continue merging the subgraph until all subgraphs are merged into the entire graph, 
except when $\Ksrc$ is fixed, in which case we stop when $\Ksrc$ segments remain. In this way, we obtain a sequence of nested segmentations, $\{\widehat\Sroi(K), K=1,\ldots, n\}$, each with a BIC value. Finally, we set $\widehat\Ksrc$ and $\widehat\Sroi$ to the values of the nested level with the smallest BIC.

At each iteration, we use an updating formula to speed the computation of BIC.
Consider the graph segmentation associated with $\Ksrc$ and the graph segmentation after merging two of the subgraphs of $\Sroi(\Ksrc)$ and label the merged subgraphs $i$ and $j$, with $1\leq i <j \leq \Ksrc$.
This merger decreases BIC by 
\begin{eqnarray}
    \Delta \mbox{BIC}_{\Ksrc, i, j} &=& 
\mbox{BIC}(\Ksrc)-\mbox{BIC}({\Ksrc-1}) \nonumber \\
    &=& 
    2 \ \Nroi{i} \log \frac{\Nroi{i{\cup}j}\Aroi{i}}{\Aroi{i{\cup}j}\Nroi{i}} \nonumber\\
    & &
    + 2 \ \Nroi{j}\log \frac{\Nroi{i{\cup}j}\Aroi{j}}{\Aroi{i{\cup}j}\Nroi{j}} \nonumber\\
     & & + \mpar\log n \,,
\end{eqnarray}
where $i{\cup}j$ denotes the union of photons that belong to subgraphs $i$ and $j$. The complete procedure for \gsrg\ is summarized in Algorithm \ref{GSRG_algo}. 

\begin{figure*}[htb!]
    \centering

    \vspace{-0.1in}
        \subfigure{
        \label{fig:illustration_circular}
        \includegraphics[width=0.8\textwidth]{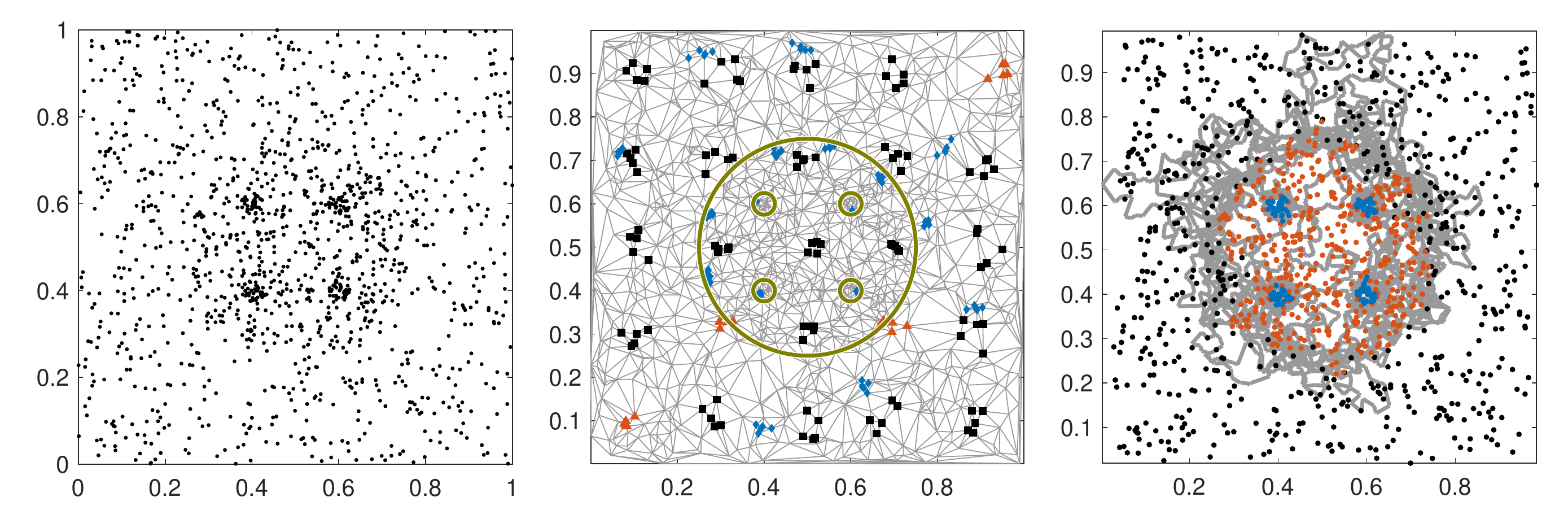}}

    \vspace{-0.15in}
        
    \subfigure{
        \label{fig:illustration_z}
        \includegraphics[width=0.8\textwidth]{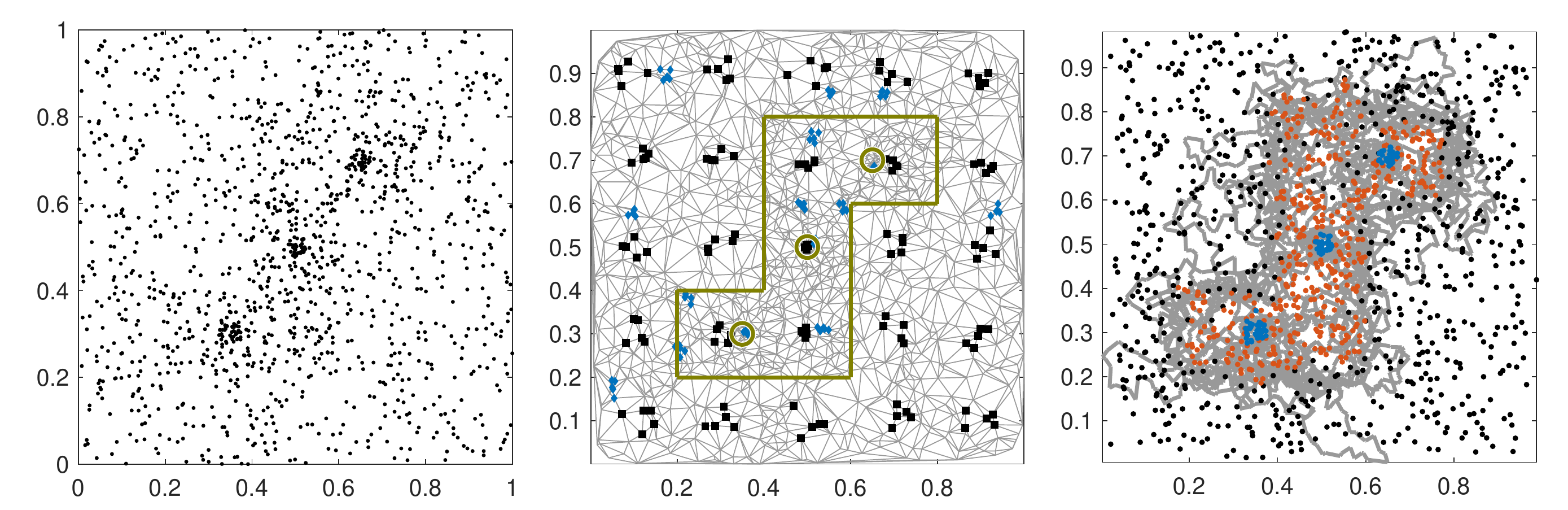}
    }
    
    \vspace{-0.15in}
    
    \subfigure{
        \label{fig:illustration_arc}
        \includegraphics[width=0.8\textwidth]{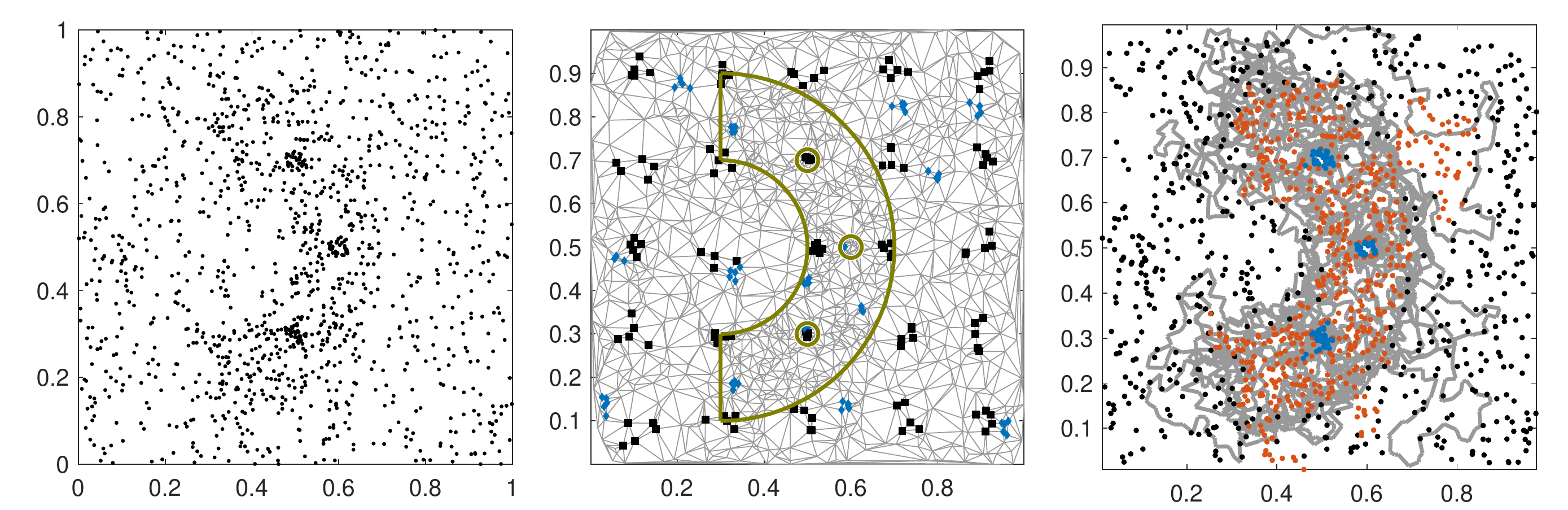}
    }
    \caption{\updatebf{Simulation studies with various shapes of the extended source (with $\parasz=1$ and $\parasnr=30$; see Section~\ref{sec:simulation}). Each row corresponds to one of the three image shapes (circular at {\sl top}, polygonal zig-zag at {\sl middle}, and semi-annular at {\sl bottom}) considered.
    \textit{Left Column:} Simulated photon locations for one simulation instance.
    \textit{Middle Column:} The graph for the exemplar (first) simulation of the left column is shown (grey lines), and the true segment boundary is overlaid (solid green lines).  The initial seeds are marked for the $5{\times}5$ regular grid (black squares), local maxima (blue diamonds), and rejected seeds (red triangles).
    \textit{Right Column:} Segment boundaries from 10 additional simulations (grey solid lines) are overlaid on the photons from the first simulation, which are marked as red for the extended source, and blue for the point-like sources.
    }
    }
    \label{fig:illustration_all_scenario}
\end{figure*}

\begin{figure}
    \centering
    \includegraphics[width=\linewidth]{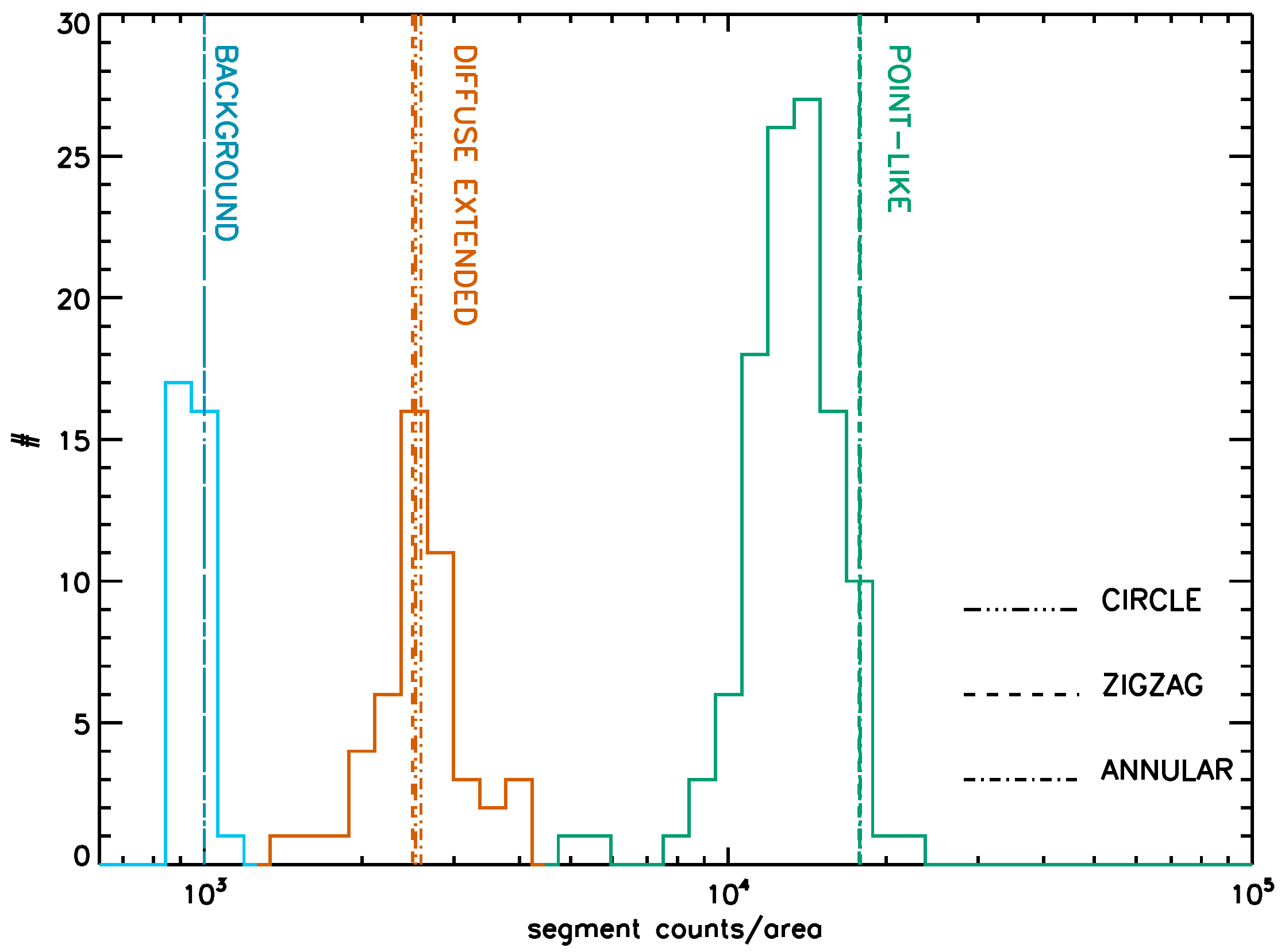}
    \caption{\updatebf{Demonstration of recovery of simulation parameters in segmentation.  The brightness of each segment recovered by \gsrg\ for all of the simulations shown in Figure~\ref{fig:illustration_all_scenario} is shown as the black histogram.  The expected brightness for each of the components (background, extended, point-like) and shapes of the extended source (circle, polygonal zigzag, and semi-annular) are marked as vertical lines with different line styles as labeled.  The background and extended emission brightnesses are recovered well, but because of the large contrast, the brightness of point-like objects shows a bias (see text).
    }}
    \label{fig:brightness_distrib}
\end{figure}

\begin{figure*}[htb!]
    \centering

    \includegraphics[width=0.48\textwidth]{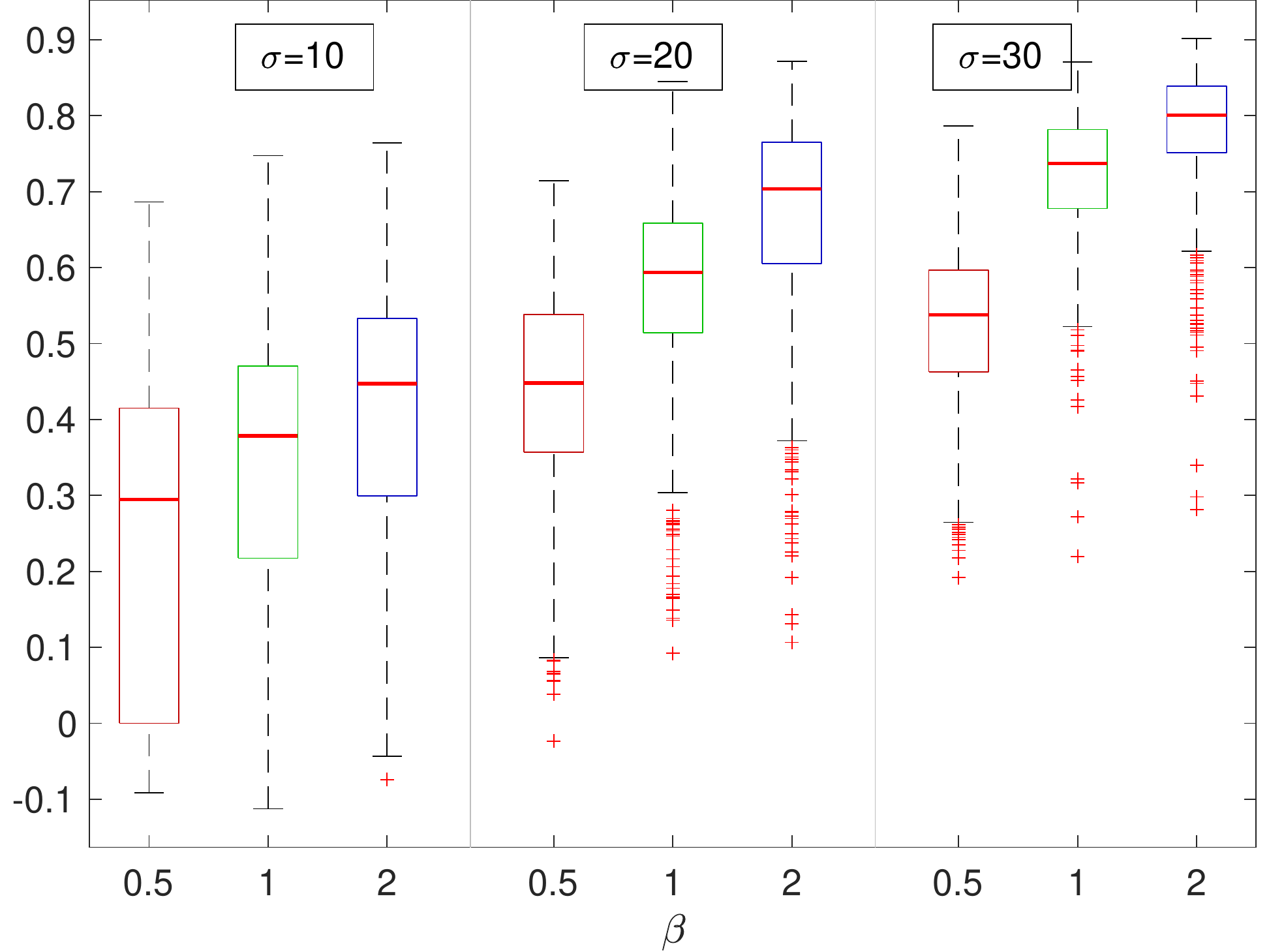} \quad
    \includegraphics[width=0.48\textwidth]{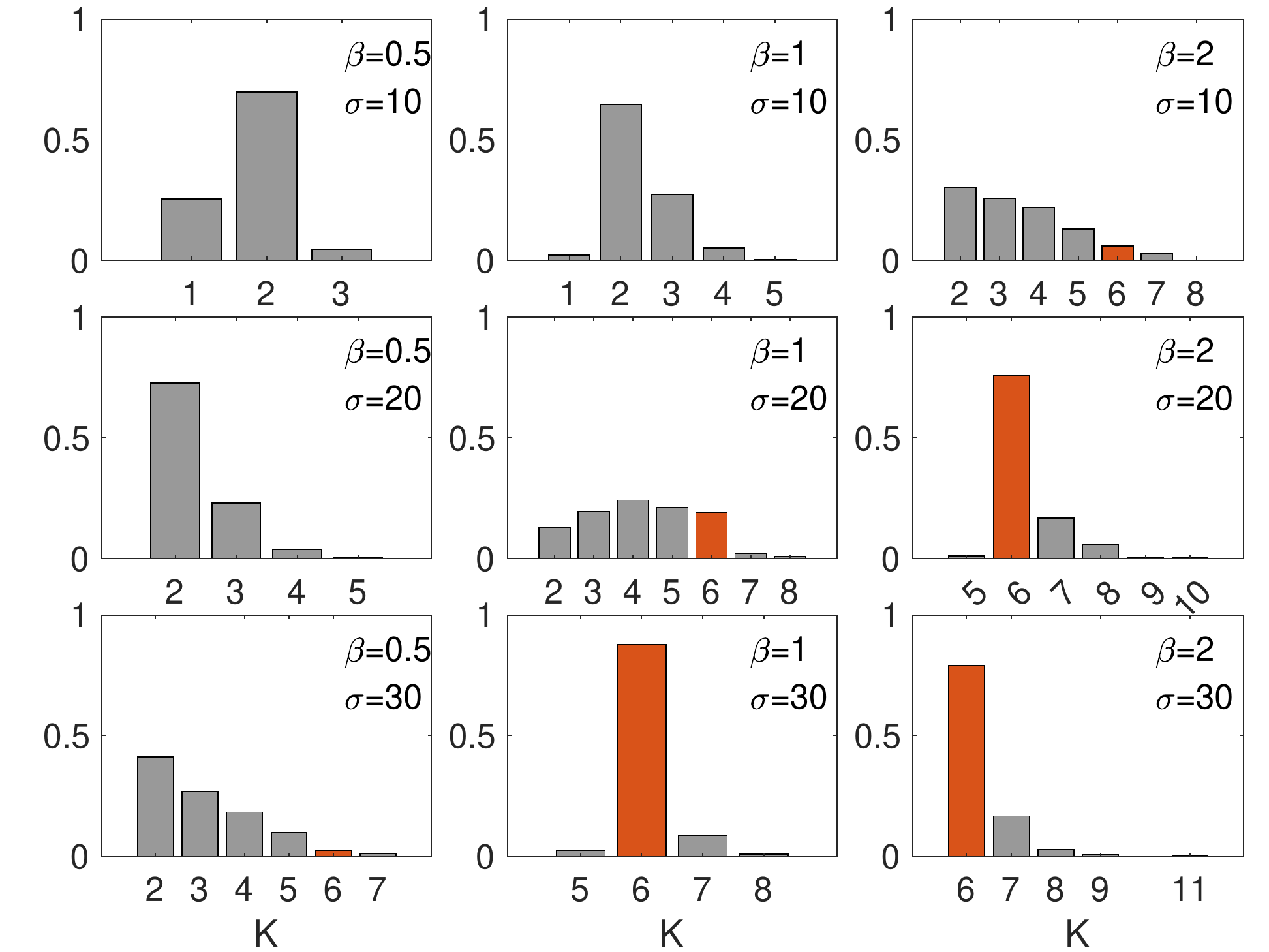}%
    
    \phantom{m}

    \includegraphics[width=0.48\textwidth]{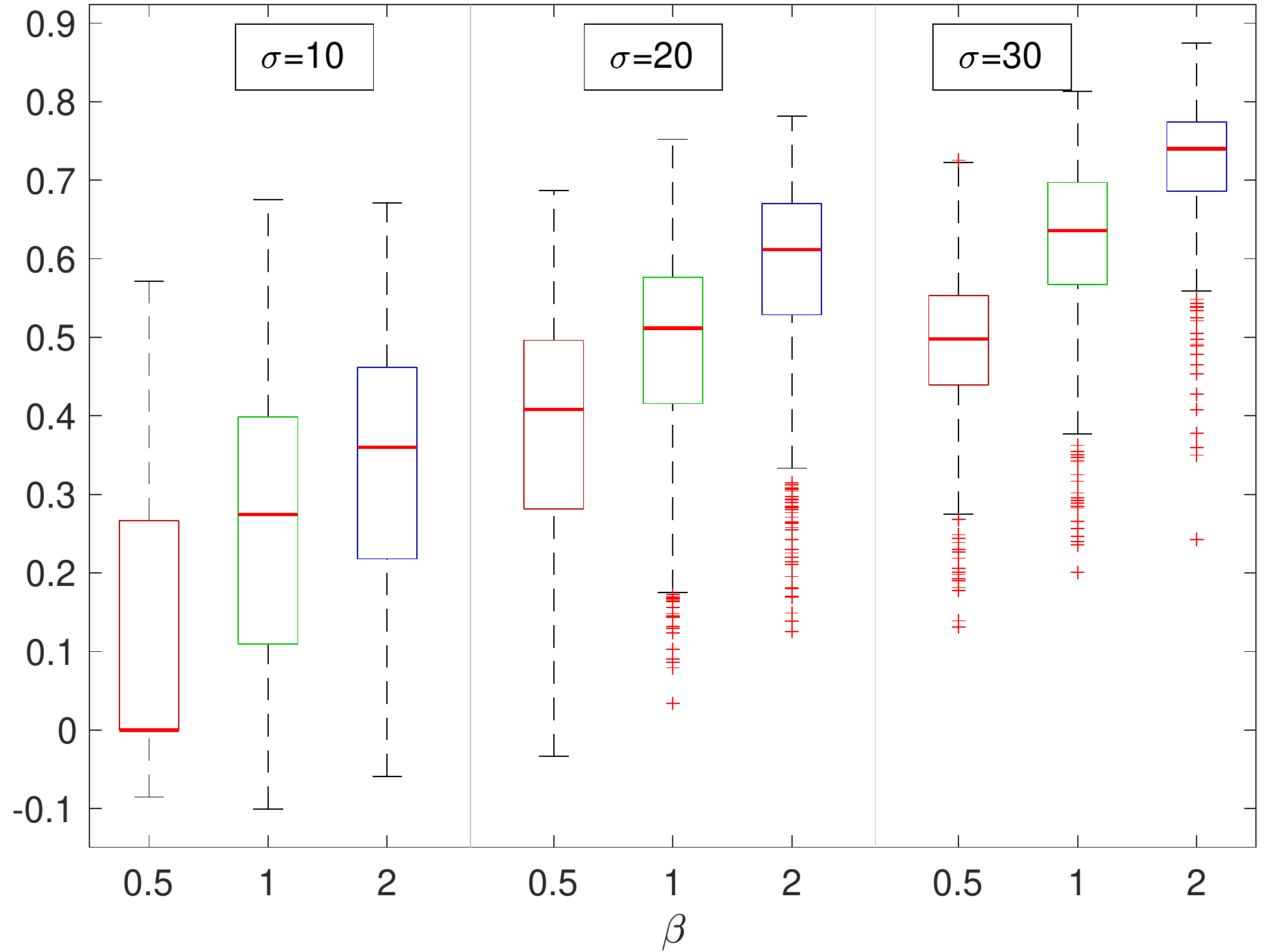} \quad
    \includegraphics[width=0.48\textwidth]{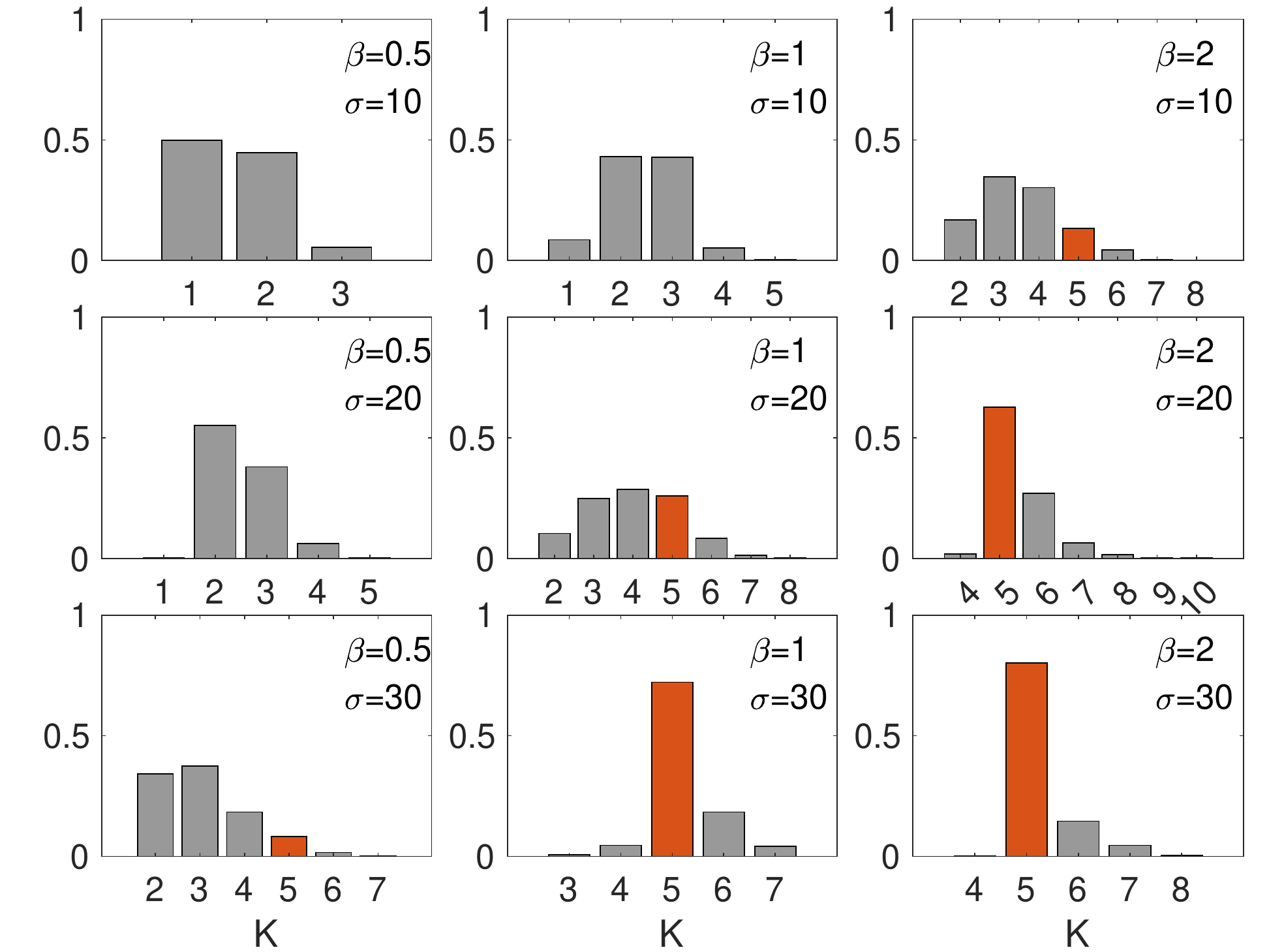}%

    \phantom{m}
      
    \includegraphics[width=0.48\textwidth]{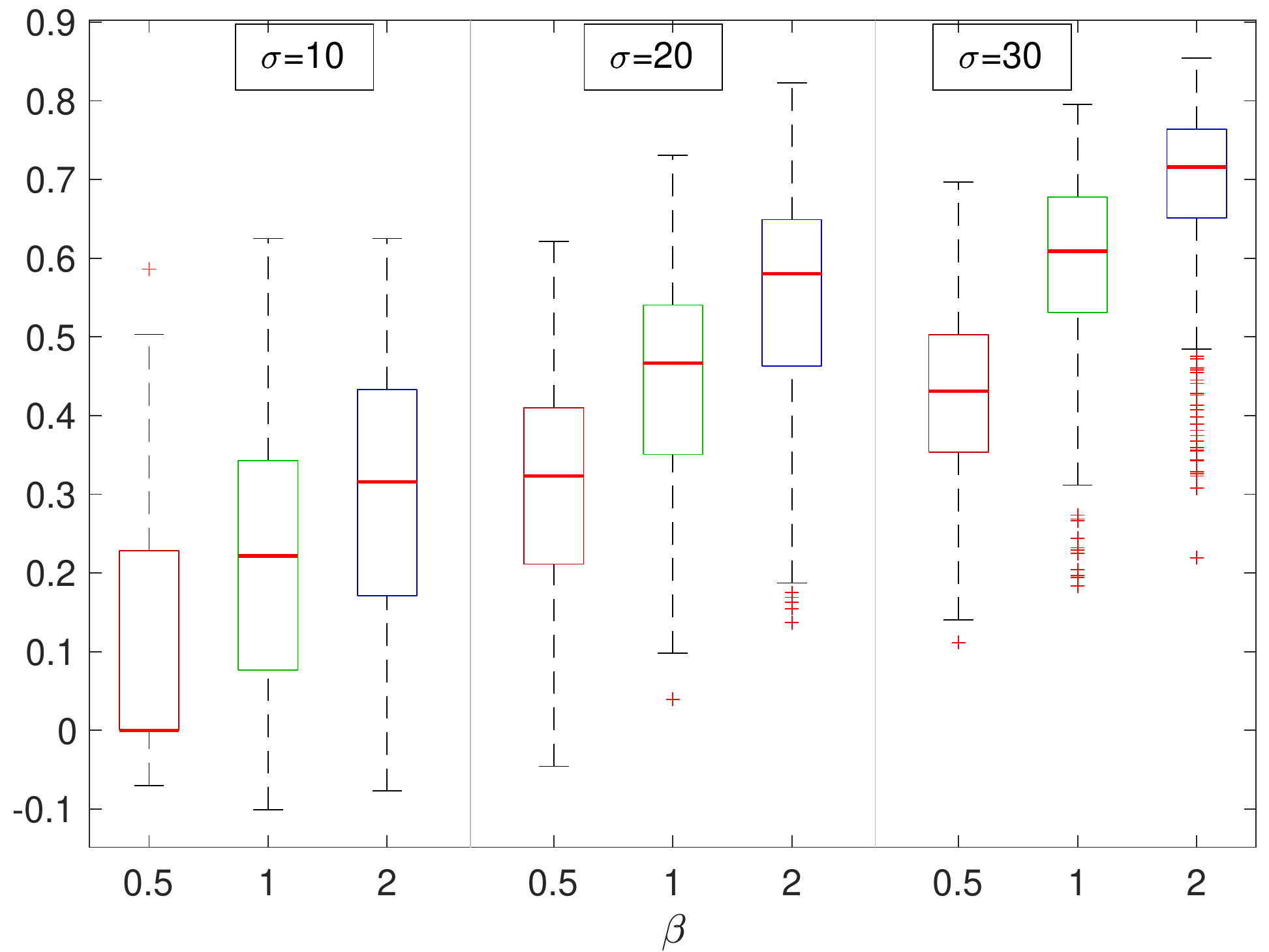} \quad
    \includegraphics[width=0.48\textwidth]{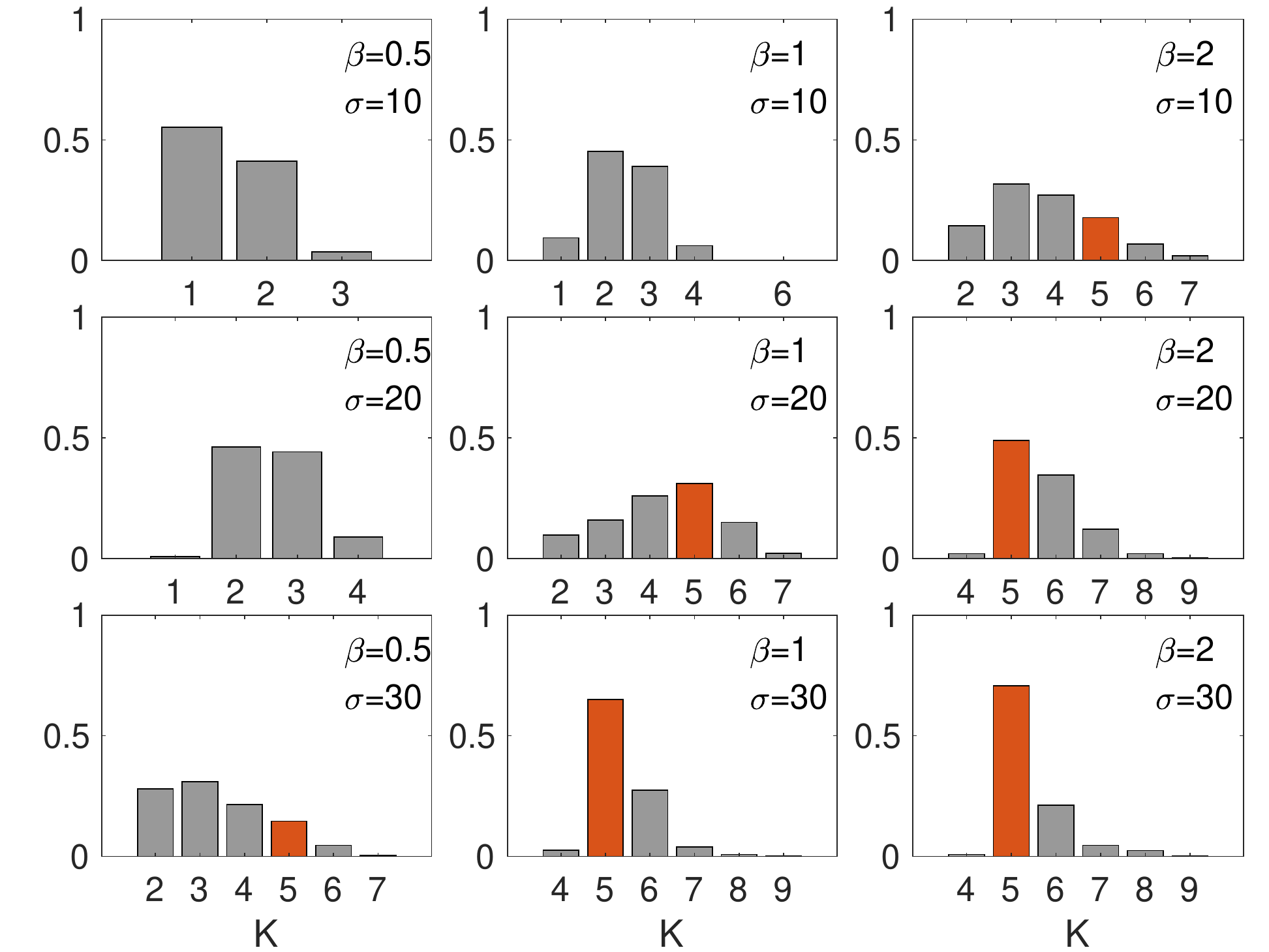}%
    \caption{Simulation studies for the various shapes of the extended source, different exposure times, $\parasz$, and different contrast ratios, $\parasnr$.  \updatebf{Each point source has $\parasz\cdot\parasnr$ counts in a circle of radius 0.025, each diffuse extended source has $10\cdot\parasz\cdot\parasnr$ counts spread uniformly over an area $\approx$0.2, and there are $1000\cdot\parasz$ counts in the background, spread uniformly across $\fov$.} The three rows correspond to the three extended source shapes in the rows of Figure~\ref{fig:illustration_all_scenario}.
    \textit{Left Column:} Boxplots of the adjusted Rand indices for $500$ replicates under each simulation setting; 
    \textit{Right Column:} Histograms 
    of the fitted number of sources $\widehat\Ksrc$ for the 500 replicates under each simulation setting,
    where the true number of sources is highlighted in red.
}
    \label{fig:metrics_all_scenario}
\end{figure*}

\section{Simulation studies}
\label{sec:simulation}

Our simulation study is conducted assuming a hypothetical instrument that produces fields of view, $\fov$,  with two-dimensional coordinates on the unit square.
The simulations are designed to assess the performance of \gsrg\ when applied to fields of view of point-like sources embedded in extended sources of different shapes, while varying the exposure time (or equivalently, the overall counts in the field) and the contrast between the different components. In all our simulation settings, ``point-like sources'' are circular sources of \updatebf{radius 0.025 and extended sources are of area $\approx$0.2 relative to the $\fov$.} 
We consider three ``true images'': \updatebf{
(a) four point-like sources embedded within a circular extended source of radius 0.25 (covering an area $0.196$ of the unit square), 
(b) three point-like sources embedded within a polygonal zig-zag shaped extended source comprising five squares of size $0.2{\times}0.2$ (total area of $0.2$), and
(c) three point-like sources embedded within an arc-shaped extended source (a half-annular shape with inner radius 0.2 and outer radius 0.4, and total area $0.189$).}
We consider these three settings because point-like sources embedded within a complex extended source are commonly observed in astrophysical fields of view, as illustrated in Section~\ref{sec:application} \updatebf{and the extended sources mimic typical astronomical shapes.  Furthermore, the variety of shapes and the contrasts considered are a stringent test of the algorithm.}  Letting $\parasz$ denote the exposure time (in arbitrary units) and $\parasnr$ denote the contrast \updatebf{level between the different components,} for each simulated $\fov$ we generated $\parasz \parasnr$ counts for each point-like source, $10\parasz \parasnr$ counts for the extended source, and $1000\parasz$ counts in expectation for the background, with the photons corresponding to each component distributed uniformly over the area allocated to it.  In Figure~\ref{fig:illustration_all_scenario}, we have adopted $\parasz=1$ and $\parasnr=30$, so \updatebf{in all cases, the point-like sources have 30~photons, the extended shapes have 300~photons, and the background has $\sim$Poisson(1000)~photons, all distributed uniformly over their allocated areas, with $\approx$200 background counts under the area of the extended source.  The contrast in surface brightness between the extended source and the background is thus $\approx$1.5$\times$, which is sufficiently large on the scale of the extended sources that the presence of the extended sources are clearly recognizable.  However, it is clear from inspection of Figure~\ref{fig:illustration_all_scenario} that local fluctuations can be sufficiently large as to make estimating the boundary of the extended sources challenging.

The photons randomly generated for each of the settings are shown for one case in the left column of Figure~\ref{fig:illustration_all_scenario}.  The middle column shows shapes of the extended sources are also shown overlaid on the corresponding Delauney triangulation for each of the photons, as well as the seeds chosen for that case.  The right column shows the segmentation, with points colored blue and extended source colored red, for the simulation in the left column, and superposed in grey lines, the result of segmentations from 10 additional simulations.  The superpositions of the segment boundary lines over the expected lines of the shapes of both the point-like and the extended sources show that while fluctuations are present in individual simulations, on average the boundaries are picked out well.  A detailed examination of the locations of the boundaries and their uncertainties requires modeling the boundaries, and we defer discussion to follow-up work (J.\ Wang, et al., in preparation).  Here, we demonstrate that the components are well recovered in all cases.  We show the distribution of the segment brightnesses found for all the simulations for all three cases in Figure~\ref{fig:brightness_distrib}: the components are clearly separated, with uncertainties of $\approx$10-15\% on the expected brightness in each component.  We find that the brightness of the point-like component suffers from a bias because of the tendency of the segment areas to preferentially encroach on the much larger area of the surrounding extended source, thus causing a downward shift in the estimated brightness.}

In our simulation design, in addition to varying the three ``true images'',  we also vary the exposure time with 
$\parasz$ taking values 0.5, 1 and 2, and the 
contrast
with $\parasnr$ taking values 10, 20 and 30. We simulate 500 fields of view under each of the 27 resulting simulation settings\footnote{The number of source counts was held fixed in all simulations, while the number of background counts was generated as a Poisson with mean $1000\parasz$ in order to explore the effect of background fluctuations.  Thus, the total counts in a given dataset is $\parasz \parasnr + 10 \parasz \parasnr + {\rm Poisson}(1000\parasz)$.}. Each of the 13,500 simulated fields of view is analyzed with \gsrg, with initial seeds specified following the ``grid supplemented by local maxima'' method of Section~\ref{sec:seed_spec}. The regular grid used for seed specification is 5-by-5, with seed size {$\numsubgraph=5$}
and a neighborhood size of $k=50$ for finding local maxima. 
Since the sources we are considering are simple, we set the BIC parameter to be $\mpar=4$;
when more complicated shapes are expected, larger values of $\mpar$
should be used. 

The second and third columns of Figure~\ref{fig:illustration_all_scenario} show the initial seed specifications and segmentation results for the first of the 500 fields of view generated with $\parasz=1$ and $\parasnr=30$. 
All the point-like sources are clearly identified.
The fitted boundaries of the extended sources are generally
quite good,  except for some mild leakage for the arc-shaped extended source. In general, we expect sources with longer perimeters per unit area\footnote{A standard measure of shape irregularity is the ``perimeter index'' of \citet{angel2010ten} which is defined to be the perimeter of a circle of area equal to that of the shape divided by the actual perimeter of the shape.} to be more challenging. This is because photons nearer the boundary of a segment are more likely to be misclassified than are those nearer the middle. Thus, more irregularly shaped sources, such as the arc-shaped source in this simulation, are more challenging, including for \gsrg.

Furthermore, several of 
the initial seeds placed by the regular grid happen to fall near the boundary of arc-shaped source, which can also jeopardize the performance of seed-based methods.

We use a clustering verification metric, specifically the \emph{adjusted Rand index} \citep[ARI;][]{Hubert-85}, to assess the quality of the \gsrg\ segmentations.
The Rand Index \citep[RI;][]{rand:71} quantifies how well a given segmentation matches the ground truth segmentation.
Specifically, each pair of photons is classified as either (a) being in the same fitted segment {\it and} in the same ground truth segment, (b) being in different fitted segments {\it and} in different ground truth segments, or (c) not being in class (a) or (b). (For the ground truth, the segments are the background, extended source, and each point-like source.) The Rand Index is defined to be the number of photons pairs in class (a) or (b), relative to the total number of photon pairs.
Thus, a perfect match to the ground truth results in RI=$1$.  The {\sl Adjusted RI} corrects the RI such that accidental overlaps of segments due to chance are accounted for, yielding values in the range $-1<\hbox{ARI}<+1$.

Figure~\ref{fig:metrics_all_scenario} summarizes the ARI
and the fitted value for the number of segments, $\widehat\Ksrc$
for the $500$ replicates under each of the 27 simulation setting.
For each of the three true images, as expected, the \gsrg\ segmentation improves as either the exposure time or the 
contrast between the brightness of the components
increases. 
This is seen in the progression from the top left panel to the bottom right panel of the right column of plots in Figure~\ref{fig:metrics_all_scenario}: the method fails to identify the embedded point sources when there are $\approx$5 counts in each source, but correctly identifies all components in $\geq$70\% of the cases (80\% for the circular and polygonal cases) when there are 60 counts in each point source.  Similarly, the ARI increases to close to one (i.e., perfect agreement between ground truth and \gsrg\ segmentation, where the fitted number of segments equals the true number of sources) as $\parasz$ and $\parasnr$ increase.

\section{Application to Antennae Galaxies}
\label{sec:application}

\begin{figure*}[tbh!]
    \centering
    \includegraphics[width=0.49\linewidth]{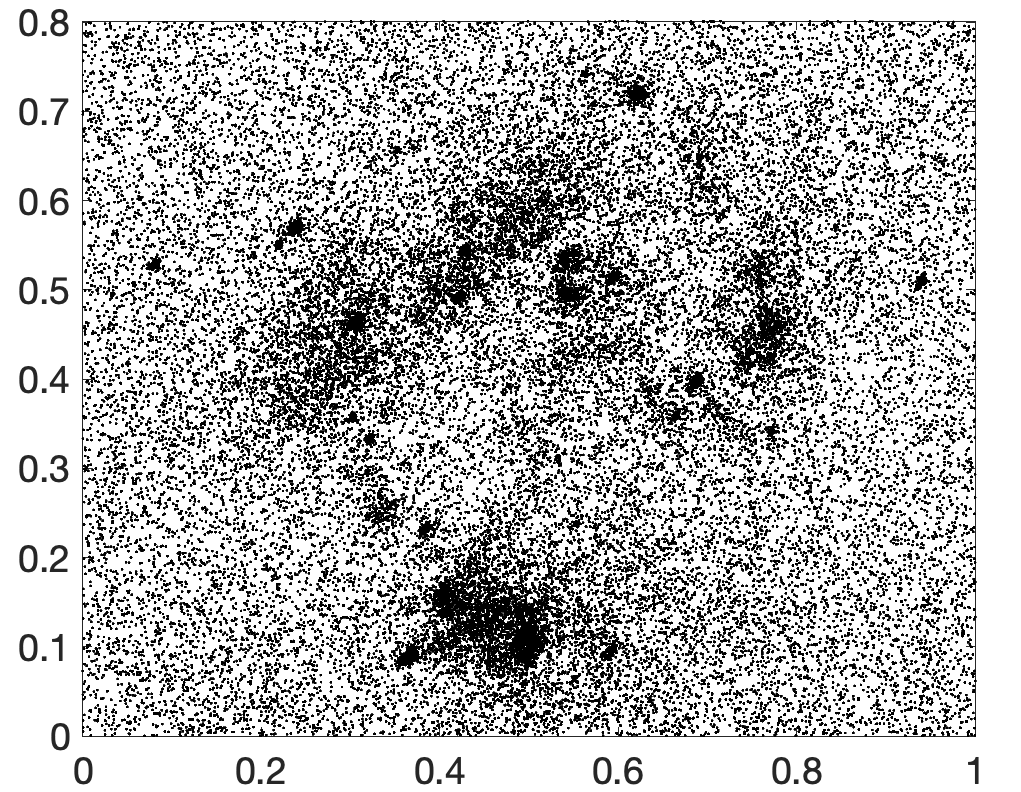}
    \includegraphics[width=0.49\linewidth]{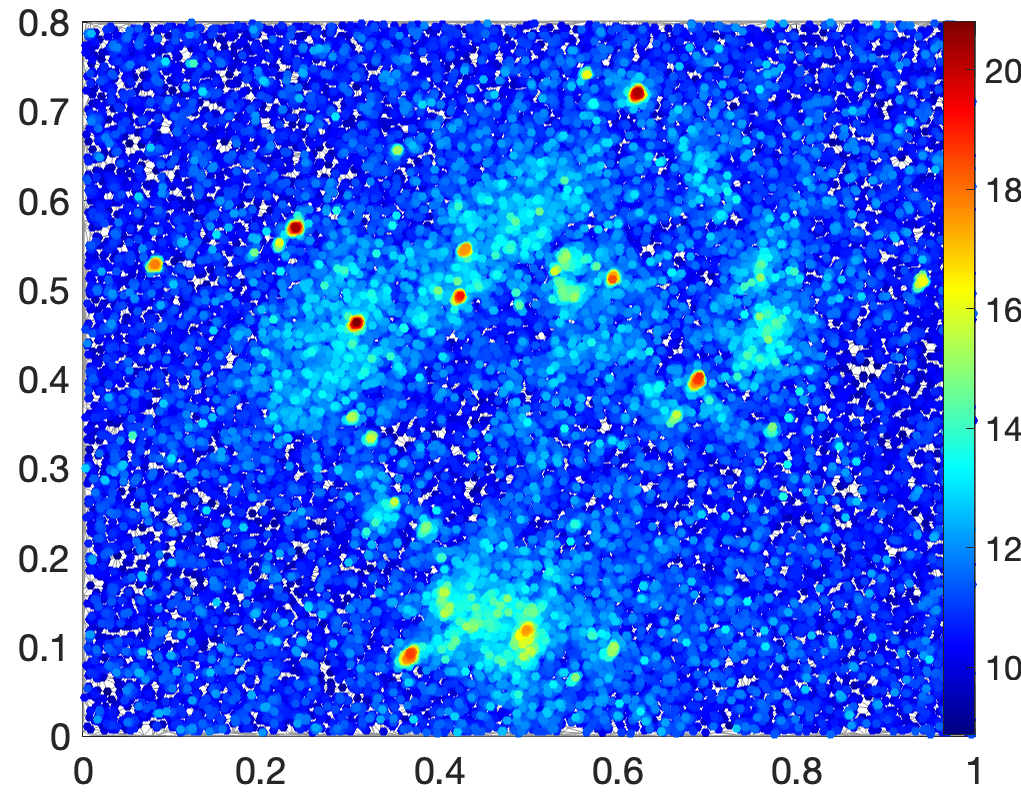}
    \includegraphics[width=0.49\linewidth]{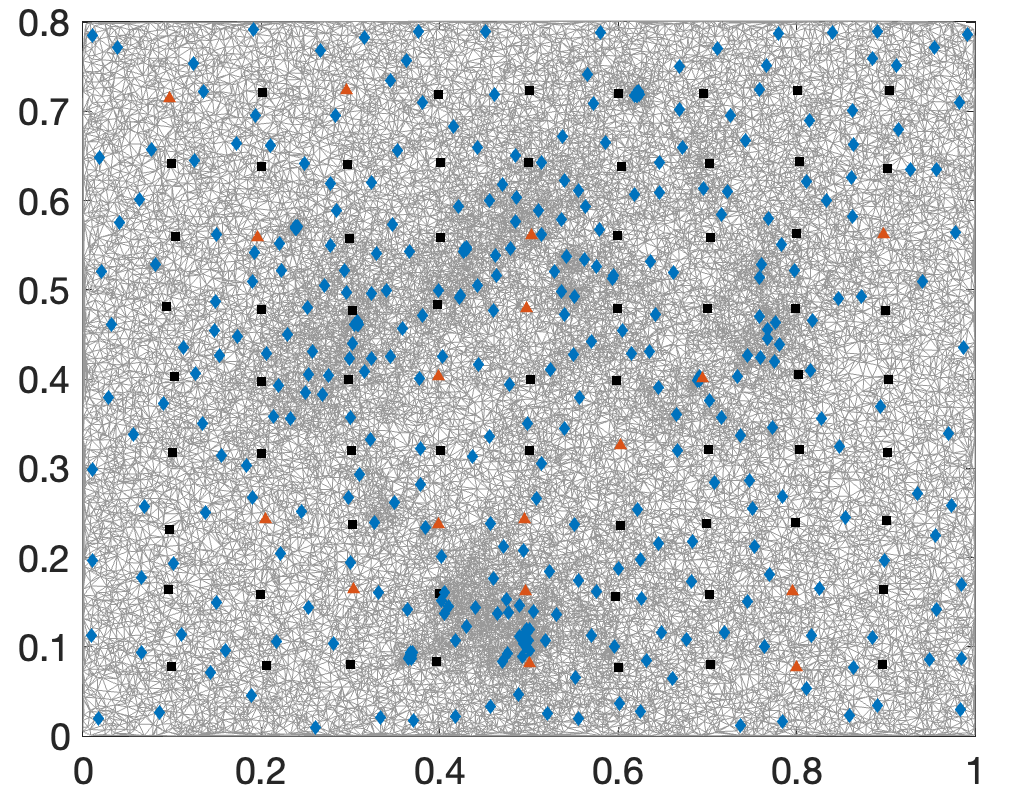}
    \includegraphics[width=0.49\linewidth]{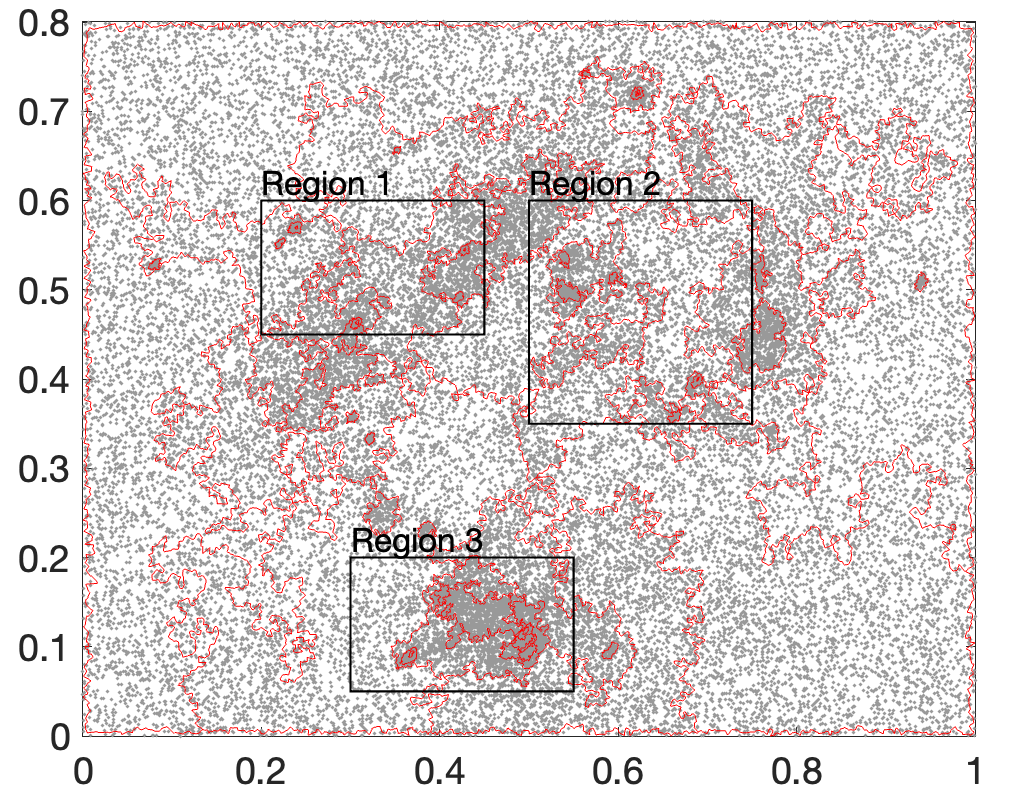}
    \caption{Representations of Antennae data.
    {\sl Top Left:} Scatterplot of the photons in from six \chandra\ observations of the Antennae galaxies carried out between December 2001 and November 2002.  The x-axis is normalized to range from 0 to 1, and the y-axis is normalized accordingly, by the same ratio, such that the data are depicted in the form that is input to \gsrg.
    {\sl Top Right:} The Delaunay triangulation of the photons is shown, with each vertex marked by a point which is color coded by the brightness determined as the reciprocal of the Voronoi area (see scale at right).
    {\sl Bottom Left:} The seeding subgraphs for \gsrg\ via a regular grid, marked with black squares, supplemented by local maxima marked with blue diamonds.  Seeds discarded due to a {strong indication}
    of being on the boundary 
    {of two segments}
    are marked as red triangles. Each subgraph is represented by one photon in the middle of the {seeding subgraph}.
    {\sl Bottom Right:} The \gsrg\ segmentation of the 
    Antennae galaxies data, with the red curves depicting the boundaries of the {fitted}
    segments. Three smaller fields of view are highlighted by black boxes, labeled Region~1-3, and magnified to show details in Figure~\ref{fig:segment_result_mag_B}.
    }
    \label{fig:antennae}
\end{figure*}

The \chandra\ observations of the Antennae galaxies provide a good test case for application of \gsrg.  The X-ray data (see Figure~\ref{fig:antennae}, top left panel) shows complex structures.
Specifically, the data reveal several point sources and extended regions, with several clumps of diffuse emission of different extent and surface brightness, along with a population of unresolved point-like sources superposed.  Some of the point sources lie within the extended sources (e.g., in the extended region at the bottom of the image) and some of the extended sources are entangled with each other.  

As a conservative scenario we used the first \chandra\ observation of the Antennae galaxies obtained on December 1st 1999 \citep[OBSID 315;][]{Fabbiano2001}. The observation was performed with the ACIS-S detector for a total exposure of  72\,ksec.
We process and screen the data \updatebf{(e.g. initial calibrations, removal of strong background flares, selection of good grades) as in \citeauthor{Zezas2002_Ant_proc} (2022; {\sl CIAO}~v3.2, CALDB~v2.11).}
 Again as a conservative scenario, we use the full dataset without any  screening for events of very low or high energies which are dominated by background.  The final dataset we use consists of 
$\approx$50,700  events within a $\sim{3.45}\arcmin\times{3.45}\arcmin$ region around the galaxy (screening for events in the generally used  0.5-8.0~keV band would result in a reduction of $\sim43\%$ in the total number of counts).

Figures~\ref{fig:antennae} show different depictions of these data, with the coordinates scaled linearly to the range $[0,1]$, 
as is assumed by our implementation of
\gsrg, and processed to show the resulting Voronoi tessellation.

We apply \gsrg\ to these data in order to obtain 
statistically meaningful non-parametric 
segmentations of the different clumps of diffuse emission, as well as to separate diffuse and point-like emission sources.  We apply the Voronoi tessellation to the photons and construct the graph of Delaunay triangulation (see top right panel of Figure~\ref{fig:antennae}). We specify the initial seeds for \gsrg\ via a regular grid supplemented by local maxima (see Section~\ref{sec:seed_spec}; shown in bottom left panel of Figure~\ref{fig:antennae}).  We start with a regular 9$\times$9 grid (i.e., $\numgrid=81$), the initial estimates of which are stabilized by assigning the $\numsubgraph=20$ nearest photons to each seed; these cover the large scale variations in the data. Local maxima are determined over a neighborhood size of $k=100$.  The 419 seeds that result from this process is a sufficiently large number to ensure that there is generally at least one seed in each point-like or extended source or the background.
Since we expect the segments of the extended sources to be more irregularly shaped in the real data than in the simulation, we choose a larger value of the BIC parameter, $\mpar=6$ (see Equation~\ref{eq:BICdefn}); this 
corresponds
to assuming that each segment has the complexity of an ellipse.

\begin{figure*}[thb!]
    \centering
    \includegraphics[width=0.48\linewidth]{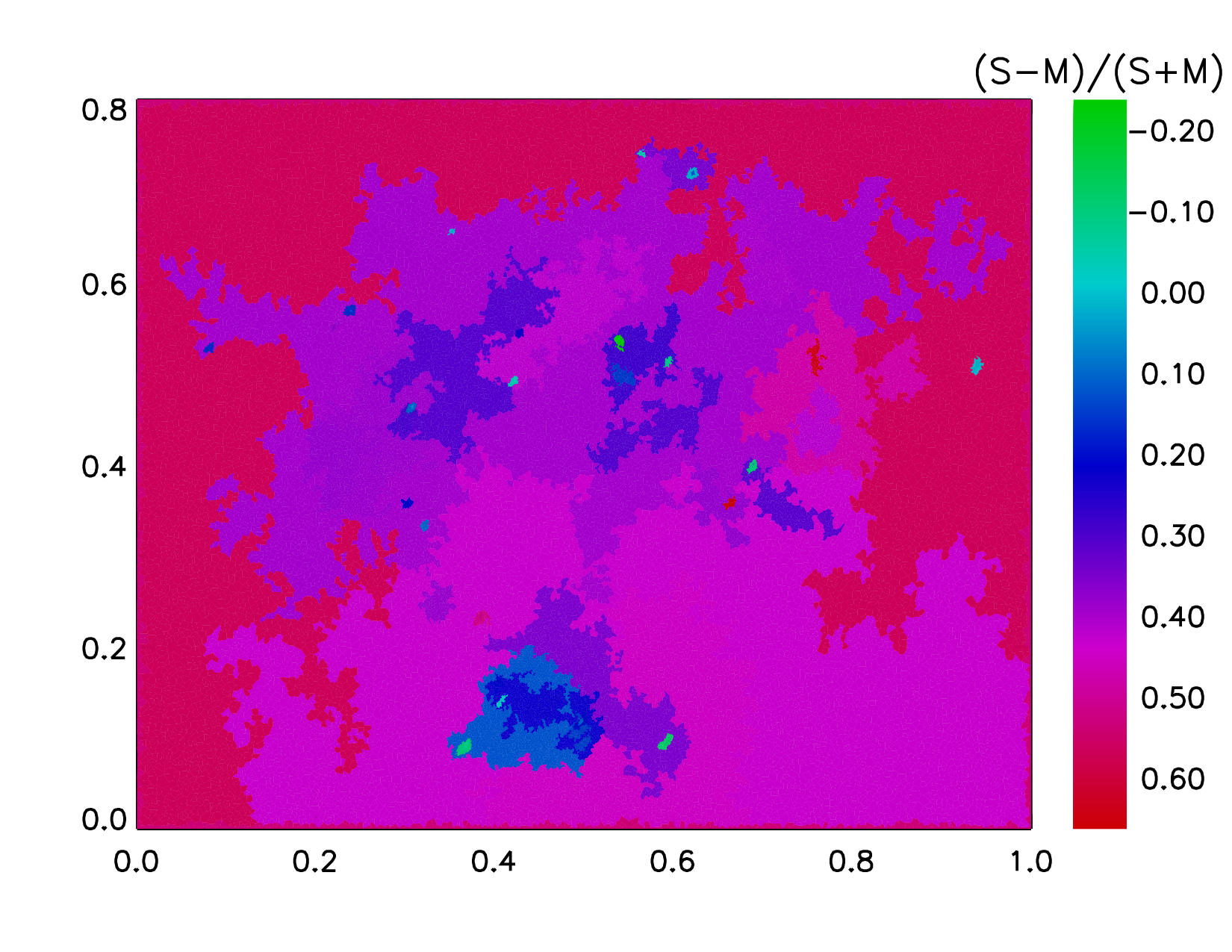}
    \includegraphics[width=0.48\linewidth]{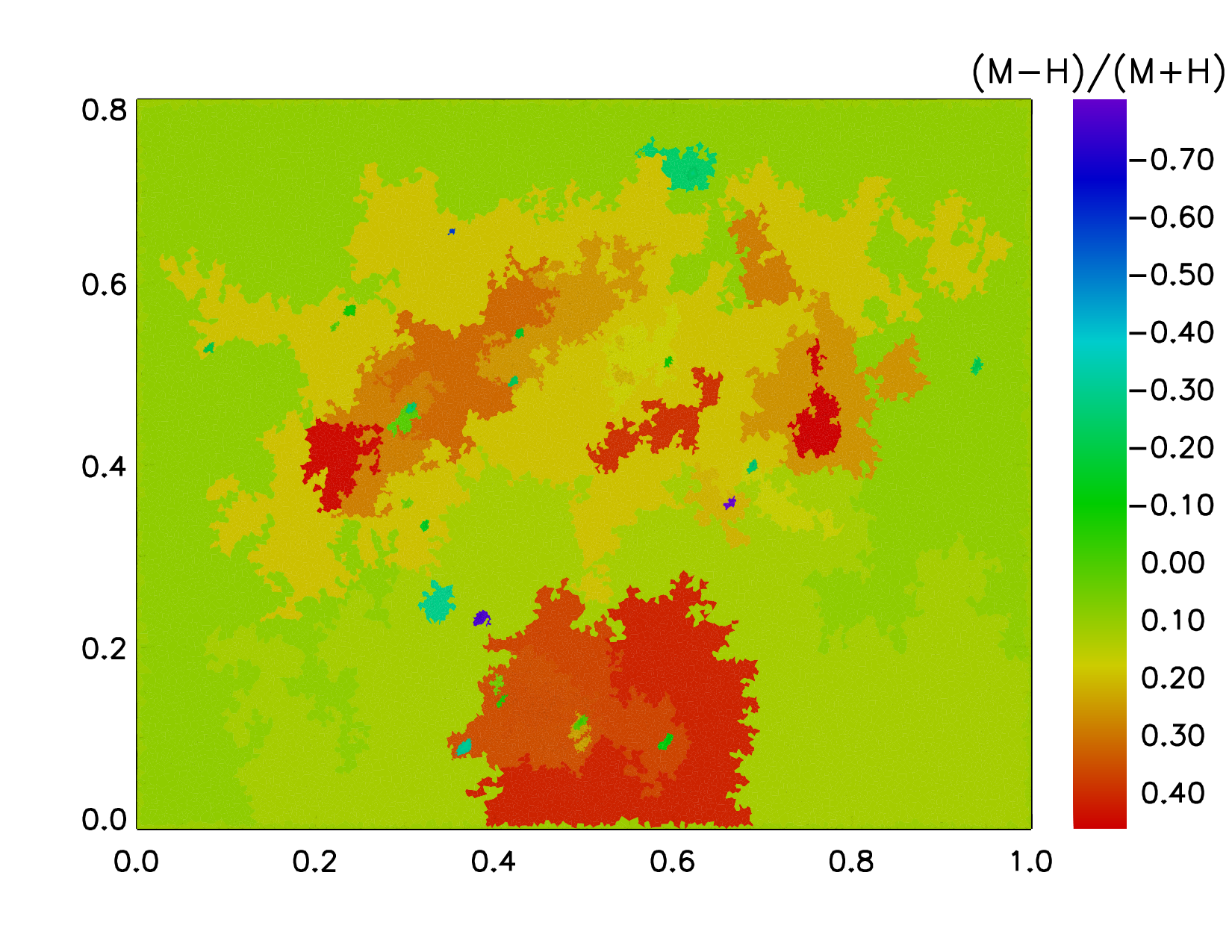}
    \caption{Fractional hardness ratios of counts in {each of the \gsrg-determined segments of the } 
    Antennae data.  The images show HR$_{SM}=\frac{S-M}{S+M}$ {\sl (left)} and HR$_{MH}=\frac{M-H}{M+H}$ {\sl (right)}, where $S, M, H$ are counts in the passbands $0.5-0.9$~keV, $0.9-1.2$~keV, and $1.2-2$~keV, respectively.  In both figures, {redder}
    colors indicate softer spectra.  The background is rendered in reddish pink in HR$_{SM}$ and light green in HR$_{MH}$. 
    }
    \label{fig:antennae_hr}
\end{figure*}

The results of \gsrg\ are shown in the bottom right panel of Figure~\ref{fig:antennae} (the regions outlined in black are discussed in more detail in Section~\ref{sec:performance_antennae}), showing
the boundaries of the fitted segments as thin red lines around the black dots depicting the photons.  \gsrg\ correctly segments areas with similar surface brightness such that photons that correspond to these diffuse components are grouped together.  The photons that belong to each of these segments can be trivially collected together for further analysis, depending on the scientific question being explored.  For instance, in Figures~\ref{fig:antennae_hr}, we show segment-wise maps of the fractional hardness ratios HR$_{SM}=(S-M)/(S+M)$ and HR$_{MH}=(M-H)/(M+H)$, where $S, M, H$ are counts in PI channels [35:61] (${\approx}$0.5:0.9~keV), [62:82] (${\approx}$0.9:1.2~keV), and [83:135] (${\approx}$1.2:2~keV), respectively.  Notice that the maps clearly demonstrate that the diffuse emission in the Antennae generally have softer spectra than the point sources.  Maps such as these can be used to identify the extent of dust lanes in the Antennae system; e.g., the segments at $\sim$(0.4,0.22), $\sim$(0.3,0.25), and $\sim$(0.6,0.75), which are characterized by harder spectra than the surrounding segments, a characteristic of increased absorption \citep[cf.,][]{Zezas2006}.  Furthermore, notice that the southern region (around Region~3 in the bottom left panel of Figure~\ref{fig:antennae}) is surrounded by a halo of relatively soft X-ray emission, in agreement with the spectral analysis of \cite{Baldi2006} who find emission from soft $\sim$0.6~keV thermal emitting gas.

\section{Discussion}\label{sec:discuss}

\subsection{Performance on the Antennae data }\label{sec:performance_antennae}

\begin{figure}[htb!]
    \centering
    \includegraphics[width=0.49\textwidth,trim={0 {.15\textwidth} 0 {.14\textwidth}},clip]{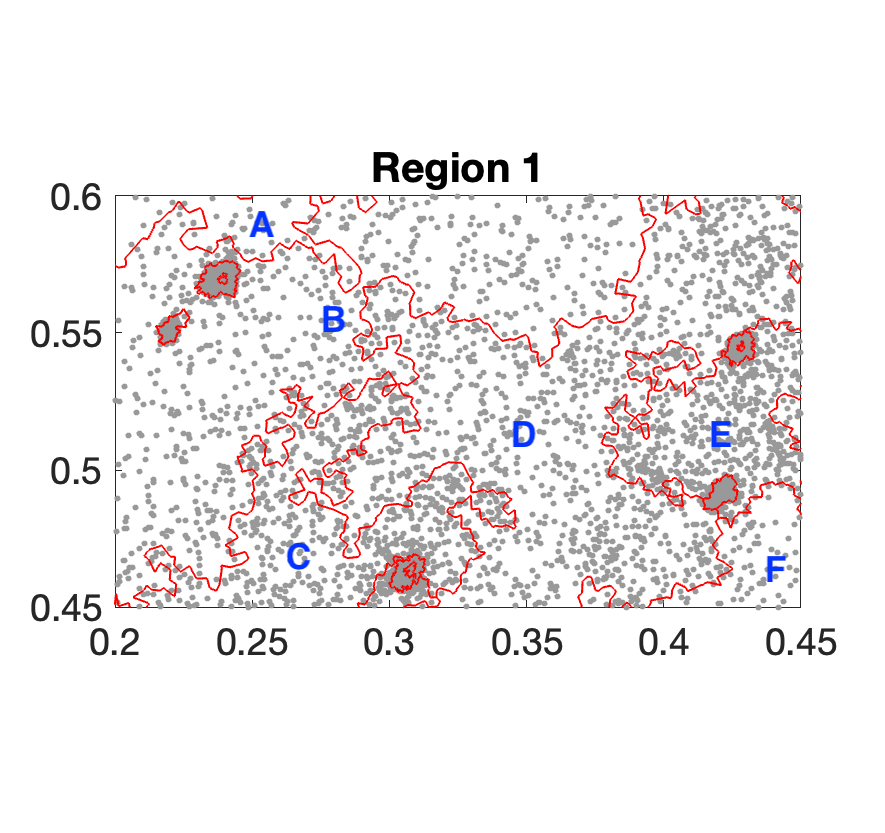} \quad
    \includegraphics[width=0.49\textwidth,trim={0 {0.02\textwidth} 0 {0.01\textwidth}},clip]{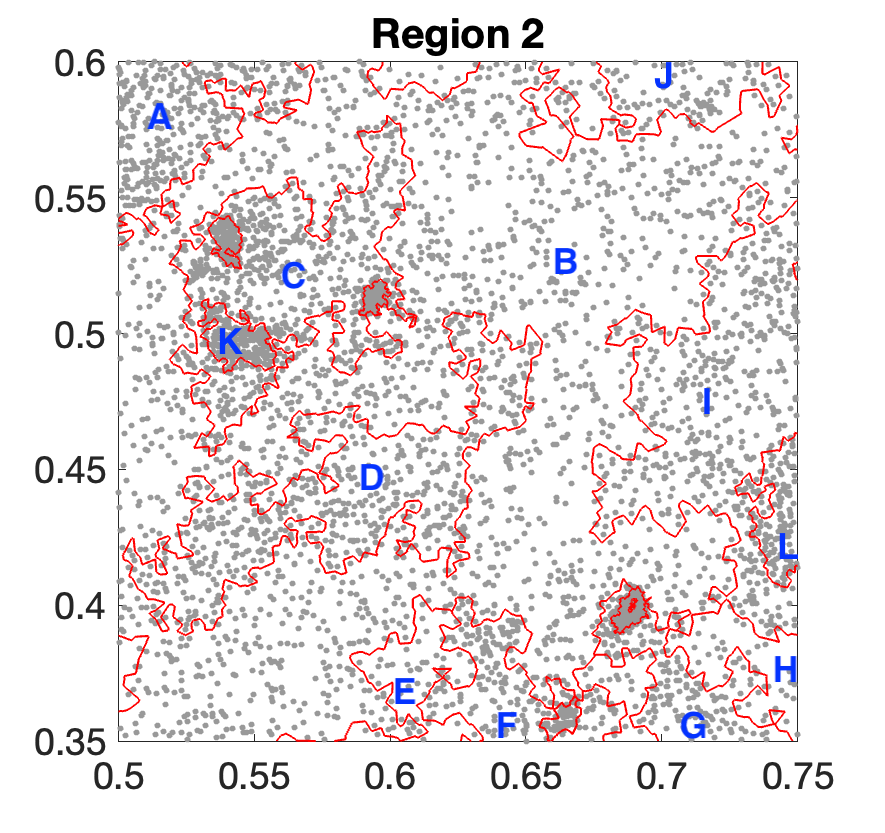} \quad
    \includegraphics[width=0.49\textwidth,trim={0 {.15\textwidth} 0 {.14\textwidth}},clip]{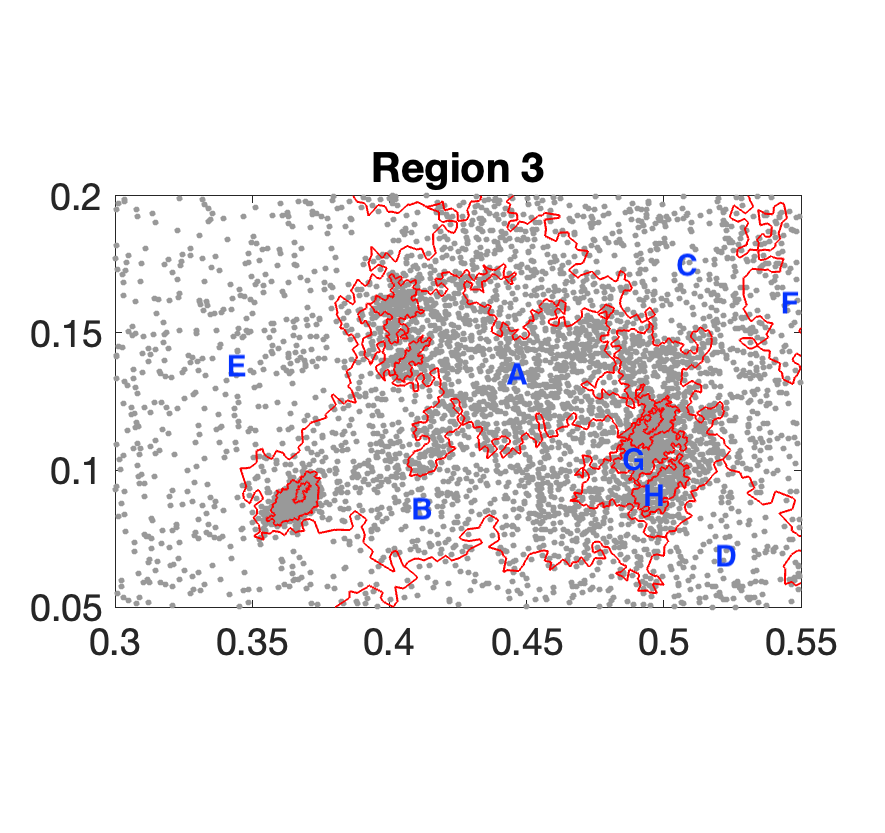}
\caption{Magnified views of the three regions of the Antennae galaxies highlighted in the bottom right panel of Figure~\ref{fig:antennae}.  The grey dots represent individual photons, and the thin red curves represent the \gsrg\ segmentation (see Section~\ref{sec:application}).  Some of the larger segments are labeled with blue letters (see text).
}
    \label{fig:segment_result_mag_B}
\end{figure}

Here we discuss the quality of the \gsrg\ segmentation of the Antennae in greater detail.  To begin with, we note that \gsrg\ successfully identifies a number of point-like sources, characterized by the presence of a large number of photons within a small space. Several of these point-like sources are superposed on extended diffuse emission and surrounded by complex structures.  Furthermore, unlike the case usually with methods that use piece-wise constant models, the point-like sources are invariably defined by single segments and not several concentric rings that approximate the typical profile of the PSF where intensity increases from the wings inward to rise to a peak at the core.  Such cases are not entirely absent, however; see 
{Figure~\ref{fig:segment_result_mag_B}, specifically}
the sources at 
$\sim$(0.23,0.57), $\sim$(0.43,0.54) and $\sim$(0.31,0.46) in Region~1,
$\sim$(0.67,0.4) in Region~2, and
$\sim$(0.49,0.12) and $\sim$(0.37,0.07) in Region~3.

Further note that the segment boundaries are not smooth because of the boundary being formed by the outermost Voronoi cells.  The photons that comprise the boundary are also subject to stochasticity, due both to PSF-induced statistical variations in photon arrival locations, as well as the greedy merging process.
Visual inspection of the results suggests that at the lowest surface brightness levels, fluctuations in the counts could result in oversegmentation of what is usually considered the background (e.g., the two large extended regions along the left side of the bottom edge of the $\fov$).
Nevertheless, the expanded views of the inset regions in Figure~\ref{fig:segment_result_mag_B} show that the segmentation correctly separates diffuse emission structures at different spatial scales and surface brightness levels.  In particular, transitions in the spatial density of photons across the boundaries are clearly discernible by eye, such as those between segments B$\leftrightarrow$C, B$\leftrightarrow$D, D$\leftrightarrow$E, D$\leftrightarrow$F, E$\leftrightarrow$F in Region~1; between segment B and segments A, C, D, F, G, I, J in Region~2; and segment A and segments B, C, D, G, and H as well as C$\leftrightarrow$F and C$\leftrightarrow$D in Region~3.  Some transitions are too subtle to be visually recognizable (e.g., A$\leftrightarrow$B in Region~1, B$\leftrightarrow$E and B$\leftrightarrow$H in Region~2, and D$\leftrightarrow$E in Region~3) but are required due to the computed contrasts in the counts per unit area.  Conversely, the brightness transitions across B$\leftrightarrow$C$\leftrightarrow$D in Region~1, B$\leftrightarrow$C$\leftrightarrow$K and B$\leftrightarrow$I$\leftrightarrow$L in Region~2, and A$\leftrightarrow$B$\leftrightarrow$C, F$\leftrightarrow$C$\leftrightarrow$D, and D$\leftrightarrow$A$\leftrightarrow$G,H are apt demonstrations of the capability of \gsrg\ to perform at the level of human visual acuity.  Parametric modeling to capture the spatial variations in such structures would be much more difficult than the segmentations achieved here.

\begin{figure*}[ht]
    \centering
    \includegraphics[width=0.485\textwidth]{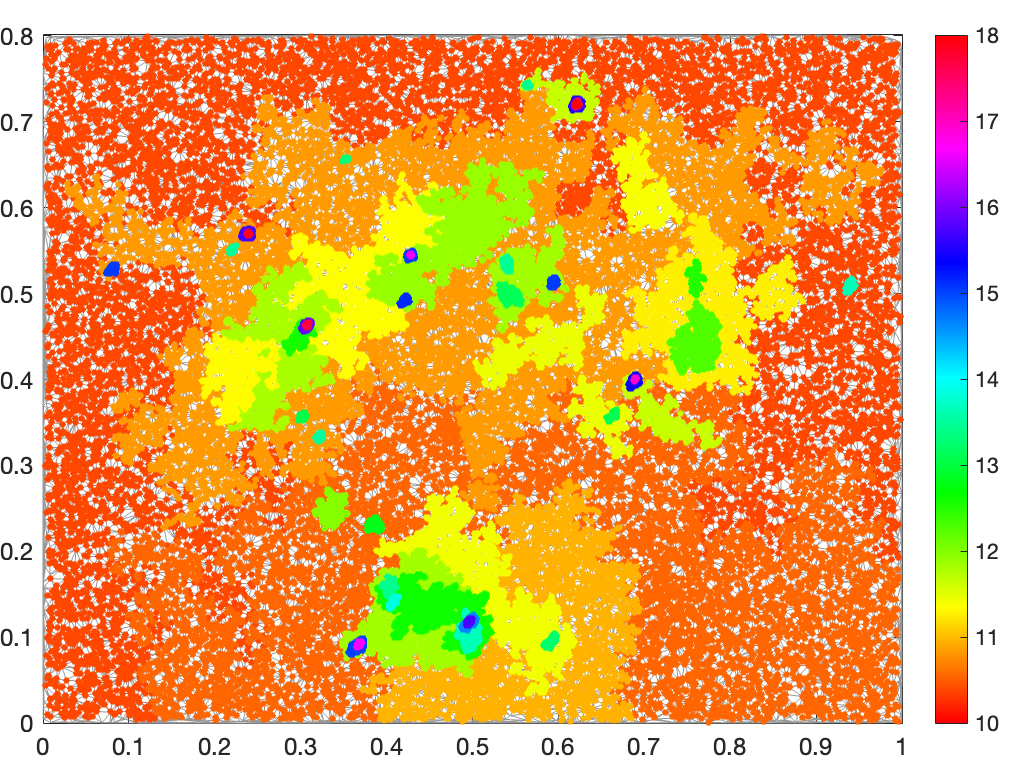}%
    \hspace{0.01\textwidth}
    \includegraphics[width=0.485\textwidth]{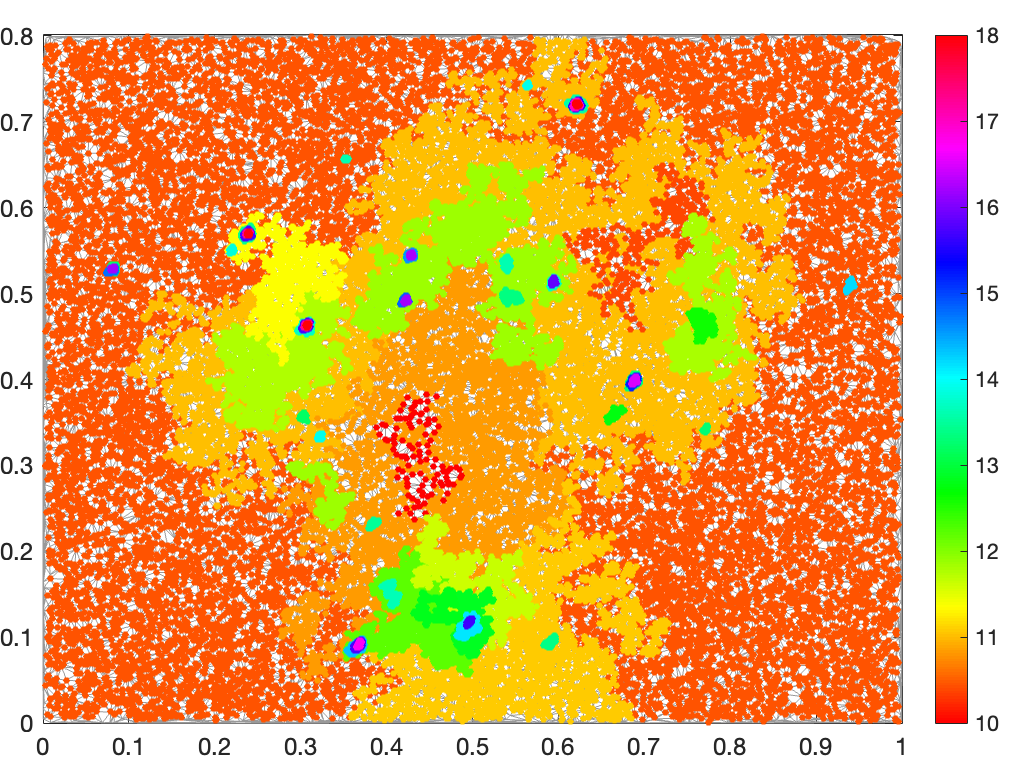}%
    \caption{Segmentation of the photon list shown in Figure~\ref{fig:antennae}. Photons are color-coded by the estimated intensity of their segment.
    \textit{Left Panel:} 55 fitted segments obtained with \gsrg\ 
    {with seeds specified via a grid supplemented by local maxima}.
    \textit{Right Panel:} 61 fitted segments obtained for brute-force segmentation, where all photons are assumed to be seeds.
    The differences between these two segmentations are described in Section~\ref{sec:performance_antennae}.
    } 
    \label{fig:segment_result}%
\end{figure*}

\begin{figure*}[ht]
    \centering
    \includegraphics[height=0.28\textheight]{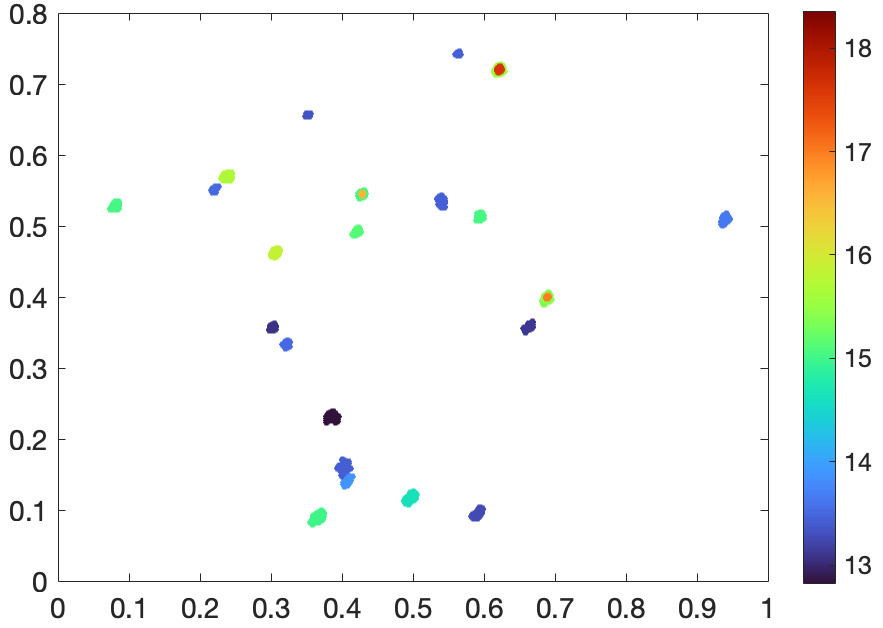}
    \hspace{0.01\textwidth}
    \includegraphics[height=0.28\textheight]{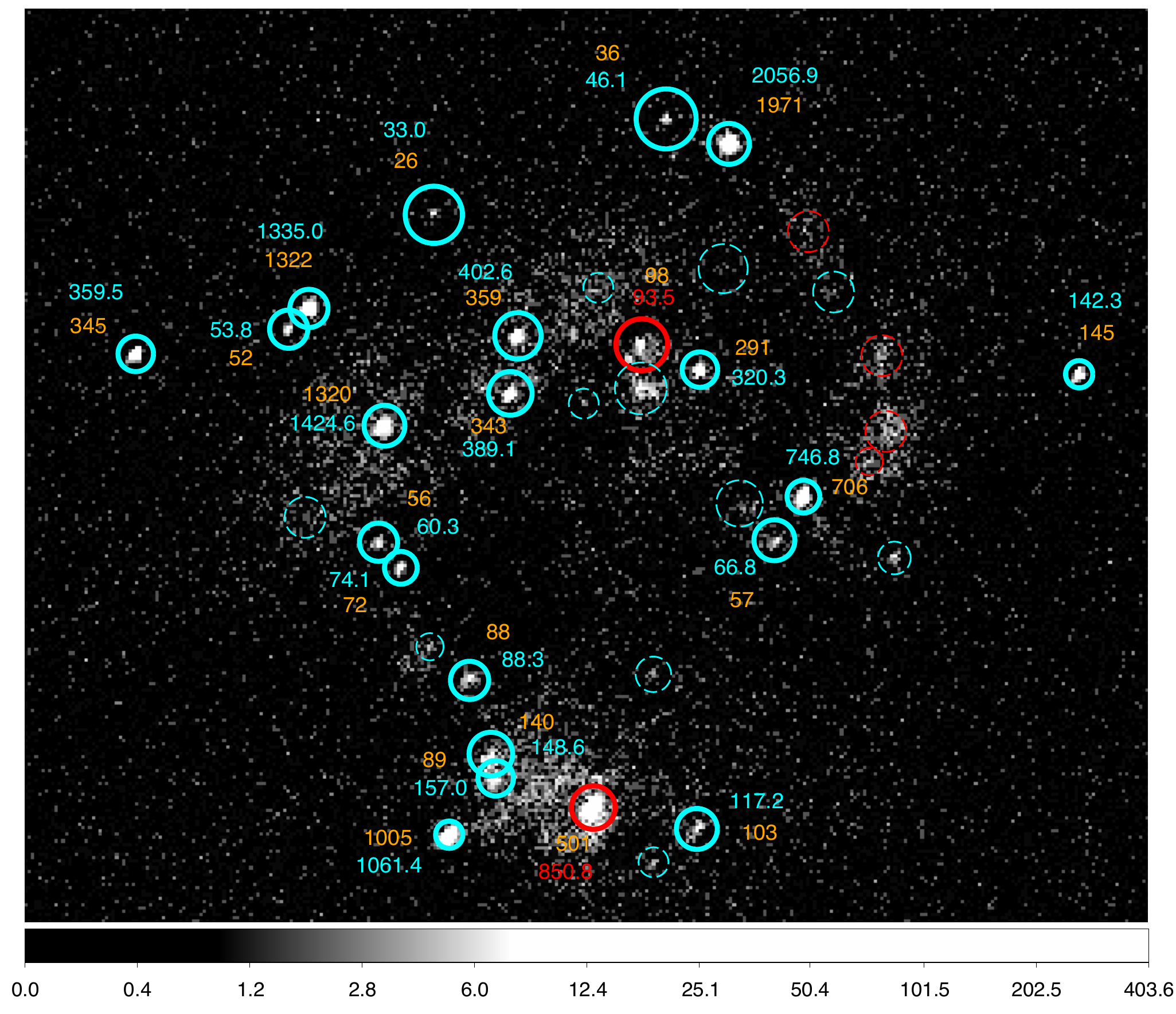}
    \caption{Identifying point-like sources in the event list data in Figure~\ref{fig:antennae}.
    \textit{Left Panel:} "Point-like" \gsrg\  segments selected to have segment area\,$\leq$0.0003 (31 total segments; 22 after concentric segments are merged).
    \textit{Right Panel:} Sources detected with {\tt wavdetect} \citep{Zezas2002_Ant_proc} overlaid on a full-band image of the \chandra\, data (OBSID 315). \updatebf{The size of the field is $2.8\times2.3$\,arcmin.} Red circles mark detected sources that were shown to be inconsistent with \chandra's PSF according to the analysis of \citep{Zezas2002_Ant_proc}, while cyan circles mark the remaining sources (some of which may also be extended). The size of the circles corresponds to the optimal {\tt wavdetect} scale.  Thicker solid circles indicate sources that are also identified by \gsrg; the remaining sources are indicated by the thinner dashed circles. The cyan and orange numbers give the source counts measured by the customized aperture photometry of \cite{Zezas2002_Ant_proc} and the \gsrg\ method respectively (the \gsrg-based counts are not background subtracted).
    Note that most strong point-like sources are identified as such by the \gsrg\ method, and in general there is good agreement in the estimated source intensities.
    }
    \label{fig:point_sources}%
\end{figure*}

An important factor in obtaining a reliable segmentation is the initial seed specification.  It is worthwhile establishing that the scheme we {propose}
generates a useful segmentation and does not miss features.  For this, we compare the \gsrg\ method against a brute-force
{segmentation}
where every photon in the dataset is taken to be a seed, and the corresponding Voronoi cells are merged
{using the BIC criteria as described in Section~\ref{sec:subgraph_merge}}. 
This brute-force scheme is similar to Scargle's (\citeyear{Scargle-02}) method (but using the BIC criteria instead of Bayes Factors) in that it eschews the
Seeded Region Growing
{on Graph} step developed and described in Section~\ref{sec:gsrg}.

In Figure~\ref{fig:segment_result}, we compare the \gsrg-based segmentation (left panel) against the brute-force segmentation (right panel).  Although at first glance the two segmentations look similar, a closer inspection reveals crucial differences.  While the quality of the identification of point-like sources does not differ significantly, there are significant differences in the diffuse emission regions that strongly favor the \gsrg\ segmentation.  Notice that segment C of Region~1 from \gsrg\ (top panel of Figure~\ref{fig:segment_result_mag_B}) is missing in the brute-force segmentation, and is effectively subsumed into segment D, which in turn also subsumes segment B.  These changes are prima facie unsupported by the visible variations in the surface density of photons.  Similarly, we see that segment E is incorrectly extended, and a different segment extends down into the middle of segment D.  Such cases are also seen in Region~2 (middle panel of Figure~\ref{fig:segment_result_mag_B}), where all of the complexity found as segments C, D, and E are lost in the brute-force segmentation; and in Region~3 (bottom panel of Figure~\ref{fig:segment_result_mag_B}) where the clear separation of segments A and B is lost in the brute-force segmentation, as is the point-like source at $\sim$(0.4,0.15).  In summary, clear variations in surface brightness are recovered in the \gsrg\ segmentation, unlike in the brute-force method.  Using a smaller set of seeds {\sl improves} the robustness of the segmentation by avoiding the chaotic development of early merging steps; errors in early stages accumulate because of the greedy merging process.  We thus conclude that \gsrg\ is superior because of, and not despite, the much
{smaller},
but perceptively selected number of seeds used to carry out the segmentation.\footnote{Just as Markov Chain Monte Carlo techniques rely on running multiple chains and verifying
{consistency}
to gain confidence in the analysis results \citep{gelm:rubi:92}, we recommend that analyses that use \gsrg\ also consider the sensitivity of the results to the adopted seed set.  The schemes that we recommend in Section~\ref{sec:seed_spec} are adequate to handle most scenarios encountered in astronomy, but are nonetheless characterized by several run-time specified parameters ($\numgrid$, $\numsubgraph$, $\numnn$, $\strata$, $\vorthr$).  Work to formalize this process via bootstrap analysis is ongoing (Jue Wu, private communication).}

Although \gsrg\ is not designed as a  point-source detection method, it is instructive to see how it behaves in the case of point-like sources.   
In Figure~\ref{fig:point_sources} we how point-like sources can be identified in \gsrg\ (left panel) compared to a wavelet-based method that is optimized to detect point sources \citep[right panel; {\tt wavdetect}; ][]{Freeman-et-al02}.  Based on the typical size of the \chandra\ point spread function (PSF), we isolate all \gsrg\ segments that cover an area of comparable or smaller size to the PSF\footnote{We choose segments identified by areas $\Aroi{k}\leq{0.0003}$ in normalized coordinates, which corresponds to a circular area of radius $\approx$1.6$''$ on the sky, comparable to the extent of the \chandra\ PSF. 
} 
and show them in the left panel. 
{(We emphasize that this is not a method to {\sl detect} point-like sources; while \gsrg\ segments with larger areas than the PSF size can be flagged as extended, regions with small areas cannot be definitively flagged as point sources, since such segments can occur due to layered segmentation of extended sources or even due to statistical fluctuations in the surface brightness of diffuse emission.)}
In the right panel, we show all the {\tt wavdetect} detected sources, superposed on a counts image of the same field.  Since {\tt wavdetect} is optimized to find point sources in a variety of scales, it may also detect  more diffuse sources. Sources that are identified as extended based on visual inspection and/or comparison with the PSF profile \citep[][e.g. lack of a core, or PSF fitting for sources with more than 100 counts;]{Zezas2002_Ant_proc}  are marked by red circles, while point-like sources are marked by cyan circles.
We note that the \gsrg\ segmentation invariably finds as ``point-like" (based on the segment area criterion) the point sources that are confirmed by the inspection process of \citet{Zezas2002_Ant_proc}  and does not find the extended sources identified by \texttt{wavdetect}. The latter are instead  components of larger diffuse emission segments. In this respect, although the \gsrg\ is not a point-source detection algorithm, screening of the identified segments based on the  segment area, $\Aroi{k}$, can be used to distinguish extended regions from point-like sources.  

\subsection{Advantages and Limitations of \gsrg}\label{sec:gsrgvgsrg}

{Our} simulations (Section~\ref{sec:simulation}) and 
{analysis of} the \chandra\ Antennae dataset (Sections~\ref{sec:application} and \ref{sec:performance_antennae}) illustrate \gsrg's strength in identifying sources at many different scales.  The method allows the identification of extended diffuse structures in X-ray data regardless of their shape, i.e., no assumptions are made about the morphology or the homogeneity of the sources.
\updatebf{Note that while the blurring due to the shape of the PSF is not explicitly modeled, this has 
negligible
effect on any source structure
at scales larger than the size scale of the PSF.  Thus,
{we expect that} useful results
{can be} obtained even when the PSF varies across the field of view, which can happen due to several reasons (the quality of the telescope optics can degrade away from the aimpoint; or fields are observed which have a large diversity of soft and hard sources, each with significantly different PSF shapes and sizes; or when complex combinations of datasets, such as multiple observations carried out at different angular offsets, are combined).}
At spatial scales larger than the PSF size, {we expect
} that results are not reliant on the specific characteristics of the PSF\footnote{Note that PSF size information {\sl may} be incorporated into the analysis, e.g., by requiring that any segment that is found to have a smaller area than that of the PSF be subsumed into a surrounding or adjacent segment.  We do not use such a criterion in this work, though such a strategy is demonstrated in Figure~\ref{fig:point_sources}.}.

\updatebf{Even when the photons are sparsely distributed, e.g., when the observation is dominated by diffuse structures at low surface brightness, and blurring due to the shape of the PSF is not included, point sources that may exist in the field of view can be identified due to the increased concentration of photons at their locations.  However, note that \gsrg\ is designed to identify large scale extended regions, so the focus and trade-offs are different.  Thus, weak point sources with low contrast against the surrounding diffuse emission are likely to be subsumed into the diffuse regions.  But because these are by definition weak, they are unlikely to contribute significantly to the brightness (or hardness) of the diffuse component.  This situation is effectively similar to the situation where the detection sensitivity of a telescope is insufficient to resolve apparently diffuse emission into its point source population.  An additional issue to consider is the bias in the point source brightness demonstrated in Figure~\ref{fig:brightness_distrib}.  This bias arises as area fluctuations from small point sources are naturally bounded at zero, but can extend into the area of the diffuse emission, leading to a skewed error distribution.  So point source intensities found by SRGonG should not be used directly, but must be re-estimated using appropriate techniques \citep[e.g.,][]{2014ApJ...796...24P}.  However, note that the bias demonstrated in Figure~\ref{fig:brightness_distrib} comes from point sources that do not have PSF wings; in real sources where point sources are sharply peaked due to the PSF, the area distribution bias works to the advantage of \gsrg, incorporating more of the PSF wings into the point source and reducing the resulting contamination of the diffuse emission by strong point sources.}

\updatebf{Unlike adaptive smoothing, source detection, or contouring methods, \gsrg\ does not set S/N thresholds or rely on thresholds of source significance to determine the presence or extent of contiguous regions.  Thus, even regions that may be characterized by low surface brightness tend not to be over-segmented.  Conversely, since the uncertainty in the estimated brightness is dependent on the number of photons that fall within the segment, small variations in adjacent regions can be more easily distinguished when the areas of the segments are sufficiently large.}

Also of note is that \gsrg\ works directly on photon lists, the most basic form of high-energy X-ray and $\gamma$-ray datasets \updatebf{and the Poisson nature of the data is explicitly accounted for during the merging process}.  While this can have detrimental effects on running time when the size of the dataset is large\footnote{For illustration, the analysis of the Antennae dataset, with $50,700$ photons and $491$ seeds, takes $\approx$240~s on a 2021 epoch 14$''$ MacBook Pro with an Apple Silicon M1 SOC.}, using the data at the highest available resolution avoids the requirements to define artificial binning sizes.

Of greater concern is the dependence of \gsrg\ results on the distribution of the initial seeds, especially for fields with low contrast. This may result in unstable behaviour because of fluctuations in the local minima in the spatial intensity distribution, leading to both false segmentation and false merging.  We caution that while the schemes we describe in Section~\ref{sec:seed_spec} are generally adequate and perform well (see, e.g., Section~\ref{sec:performance_antennae}), 
as is typical with seeded-region-growing methods, the sensitivity of the segmentation to the adopted seed structure must always be checked.

Future extensions of this method will include quantification of the  uncertainty of the segmentation resulting from the stochastic nature of the data, which would be quantified in terms of uncertainty on
{the number of segments,} the outline of the
{segments,}
and the corresponding 
source flux
{within each segment}. Other avenues to explore include 
different merging procedures as substitutes for the greedy merge to address the over-segmentation.
{The goal of }
such alternative merging options
{would be to} search more possible 
{final segmentations} and make the algorithm more robust to seed initialization.
Yet another potential extension is to perform the analysis in 3 dimensions, incorporating photon energy information.  Currently spectral information can only be used by running the code on passband filtered data.
  
\section{Summary}\label{sec:summary}

We have developed an algorithm that provides a piece-wise 
{constant} segmentation of 
a photon event list
that approximates the 
{spatial} structure present in the data.  Point-wise surface brightnesses are
initially
estimated as the inverse of their Voronoi cell areas 
and cells with similar brightnesses are grouped together to grow segments.  The seeds 
{needed to grow the segments}
can be initialized as regular grids, additionally supplemented with local maxima, or set using more complex processes by stratified sampling of Voronoi cell areas.  The process begins with a deliberate over-segmentation, and neighboring segments are sequentially merged by maximizing the BIC change. 
The resulting (greedy) segmentation generates apertures on the sky plane that can be used to collect photons and carry out further analysis in a way that removes manual intervention in selecting regions of interest.  We have explored this method via both simulations and application to a complex \chandra\ dataset, and find that it consistently provides a good description of both point-like and extended diffuse regions of arbitrary shapes.  

We note that this is not a source detection method,  but a robust method for the definition of source regions, especially for extended sources. In this way, it can be used to perform photometry or spectroscopy on arbitrarily shaped extended sources.

This method provides several advantages over other commonly used methods for the analysis of extended sources in 
high-energy photon
data. Namely, it allows the identification of sources at different scales even when they are embedded within each other without imposing any restrictive assumptions on the 
{spatial} distribution of the source photons or
{the source} intensity.    

\facilities
\chandra\ (ACIS)

\software {\sl CIAO} \citep[\url{https://cxc.harvard.edu/ciao/};][]{Fruscione-06};
{\sl PINTofALE} \citep{2000BASI...28..475K};
{\sl Matlab} (\url{https://www.mathworks.com/products/matlab.html});
\gsrg\ (\url{https://github.com/jujWang96/Astro_sim}).

\section*{Acknowledgements}
This work was conducted under the auspices of the CHASC International Astrostatistics Center. CHASC is supported by NSF DMS-18-11308, DMS-18-11083, DMS-18-11661, 
DMS-21-13615, DMS-21-13397, and DMS-21-13605; by the UK Engineering and Physical Sciences Research Council [EP/W015080/1]; 
and by NASA
18-APRA18-0019.
We thank our CHASC colleagues for many helpful discussions, especially Jilei Yang for his valuable comments on an earlier draft.
MF, JW and TCML acknowledge further support from NSF through CCF-19-34568, DMS-18-11405 and DMS-19-16125.
VLK further acknowledges support from NASA contract to the Chandra X-ray Center NAS8-03060.
DvD and AZ were also supported in part by a Marie-Skodowska-Curie RISE (H2020-MSCA-RISE-2015-691164, H2020-MSCA-RISE-2019-873089) Grants provided by the European Commission. 

\newpage
\bibliography{refs}
\bibliographystyle{aasjournal}



\appendix

\section{Nearest Neighbor Labeling} \label{sec:percolate}

Here we describe the heuristic by which selected photons are collected into groups characterized by their proximity (used in the Voronoi-area stratified sampling scheme for seed specification; see Section~\ref{sec:seed_spec}).  The photons considered in a given stratum are defined by a small range of Voronoi areas, or analogously, are located at similar contour levels if an image were constructed from the photons.  Thus, they are likely to be sparsely distributed, but with clumps of photons surrounding higher intensity regions.  The goal here is to group the clumped photons that are near each other, without breaking up rings or other complex shapes.  We emphasize that this heuristic is a quick but approximate pre-processing method to pick seeds for the full-fledged \gsrg\ algorithm.  We expect this heuristic to be useful in situations where the astronomical dataset is characterized by sparsely distributed structures with a large dynamic range in surface brightness.

We first determine an average characteristic length scale for the ensemble of photons included in 
\newcommand{\stratum}{\Upsilon}
stratum $\stratum$, as
$$
{{L}}_{\stratum}
= 2 \sqrt{\frac{1}{2}\left[\max_{i\in\stratum}\{\Avor{i}\} 
+ \min_{i\in\stratum}\{\Avor{i}\}\right]}.
$$
This ensures that the length scale is typical of stratum $\stratum$.  We begin with an arbitrary photon from $\stratum$, assigning it a unique 
group
label, and recursively assign
this group label to any neighbor, i.e., any photon in stratum $\stratum$ located within a Euclidian distance of
$L_\stratum$
from any photon assigned to this group.
The recursive labeling ends when no new neighbors are present, and we move to another arbitrary as yet unlabeled photon
in $\stratum$,
assign it a different label, and repeat the process.  We continue this labeling until all photons
in $\stratum$
are assigned labels.   For a case where the photons are placed uniformly on a regular grid, this results in all the photons being aggregated into one clump with one label.  If there are multiple clumps separated by
$>L_\stratum$,
each clump will be assigned a separate label.  We eventually discard all clumps with fewer than $\vorthr$ photons and do not use them to set a seed.
The entire process is repeated for each of the $\strata$ strata.

\end{document}